\documentclass[11pt,amsfonts]{amsart}

\newtheorem{theorem}{Theorem}[section]

\newtheorem{definition}[theorem]{Definition}

\newtheorem{remark}[theorem]{Remark}

\newtheorem{example}[theorem]{Example}

\begin{document}

\title[]{Cheeger-Gromov Theory and Applications to General Relativity}

\author[]{Michael T. Anderson}

\thanks{Partially supported by NSF Grant DMS 0072591}

\maketitle

\tableofcontents

\setcounter{section}{0}

 This paper surveys aspects of the convergence and degeneration of 
Riemannian metrics on a given manifold $M$, and some recent 
applications of this theory to general relativity. The basic point of 
view of convergence/degeneration described here originates in the work 
of Gromov, cf. [31]-[33], with important prior work of Cheeger [16], 
leading to the joint work of [18]. 

 This Cheeger-Gromov theory assumes $L^{\infty}$ bounds on the full 
curvature tensor. For reasons discussed below, we focus mainly on the 
generalizations of this theory to spaces with $L^{\infty},$ (or 
$L^{p})$ bounds on the Ricci curvature. Although versions of the 
results described hold in any dimension, for the most part we restrict 
the discussion to 3 and 4 dimensions, where stronger results hold and 
the applications to general relativity are most direct. The first three 
sections survey the theory in Riemannian geometry, while the last three 
sections discuss applications to general relativity. 

 I am grateful to many of the participants of the Carg\`ese meeting for 
their comments and suggestions, and in particular to Piotr Chru\'sciel and 
Helmut Friedrich for organizing such a fine meeting.

\section{Background: Examples and Definitions.}
\setcounter{equation}{0}

 The space ${\mathbb M}$ of Riemannian metrics on a given manifold $M$ 
is an infinite dimensional cone, (in the vector space of symmetric 
bilinear forms on $M$), and so is highly non-compact. Arbitrary 
sequences of Riemannian metrics can degenerate in very complicated 
ways. 

  On the other hand, there are two rather trivial but nevertheless 
important sources of non-compactness.

\medskip

$\bullet$ {\it Diffeomorphisms}. The group $\mathcal{D}$ of 
diffeomorphisms of $M$ is non-compact and acts properly on ${\mathbb 
M}$ by pullback. Hence, if $g$ is any metric in ${\mathbb M}$ and 
$\phi_{i}$ is any divergent sequence of diffeomorphisms, then $g_{i} = 
\phi_{i}^{*}g$ is a divergent sequence in ${\mathbb M}$, (at least if 
the manifold $M$ is compact for instance). However, all 
the metrics $g_{i}$ are isometric, and so are indistinguishable 
metrically. In terms of a local coordinate representation, the metrics 
$g_{i}$ locally are just different representatives of the fixed metric 
$g$.

 Thus, for most problems, one considers only equivalence classes of 
metrics $[g]$ in the moduli space 
$$\mathcal{M}  = {\mathbb M} /\mathcal{D}.$$
(A notable exception is the Yamabe problem, which is not well-defined 
on $\mathcal{M}$, since it is not invariant under $\mathcal{D}$).

$\bullet$ {\it Scaling}. For a given metric $g$ and parameter $\lambda  
> $ 0, let $g_{\lambda} = \lambda^{2}g$ so that all distances are 
rescaled by a factor of $\lambda .$ If $\lambda  \rightarrow  \infty ,$ 
or $\lambda  \rightarrow $ 0, the metrics $g_{\lambda}$ diverge. In the 
former case, the manifold $(M, g_{\lambda})$, say compact, becomes 
arbitrarily large, in that global invariants such as diameter, volume, 
etc. diverge to infinity; there is obviously no limit metric. In the 
latter case, $(M, g_{\lambda})$ converges, as a family of metric 
spaces, to a single point. Again, there is no limiting Riemannian 
metric on $M$.

\medskip

 Although one has divergence in both cases described above, they can be 
combined in natural ways to obtain convergence. Thus, for $g_{\lambda}$ 
as above, suppose $\lambda  \rightarrow  \infty ,$ and choose any fixed 
point $p\in M.$ For any fixed $k  > $ 0, consider the geodesic ball 
$B_{p} = B_{p}(k/\lambda ),$ so the $g$-radius of this ball is 
$k/\lambda  \rightarrow $ 0, as $\lambda  \rightarrow  \infty$. On the 
other hand, in the metric $g_{\lambda},$ the ball $B_{p}$ is a geodesic 
ball of fixed radius $k$. Since $k/\lambda $ is small, one may choose a 
local coordinate system $\mathcal{U} = \{u_{i}\}$ for $B_{p},$ with $p$ 
mapped to the origin in ${\mathbb R}^{n}.$ Let $u_{i}^{\lambda} = 
\lambda u_{i} = \phi_{\lambda} \circ u_{i}$, where $\phi_{\lambda}(x) = 
\lambda x$. Thus $\phi_{\lambda}$ is a divergent sequence of 
diffeomorphisms of $\mathbb{R}^{n}$, and $\mathcal{U}_{\lambda} = 
\{u_{i}^{\lambda}\}$ is a new collection of charts. One then easily 
sees that
\begin{equation} \label{e1.1}
g_{\lambda}(\partial /\partial u_{i}^{\lambda}, \partial /\partial 
u_{j}^{\lambda}) = g(\partial /\partial u_{i}, \partial /\partial 
u_{j}) = g_{ij}. 
\end{equation}
As $\lambda  \rightarrow  \infty ,$ the ball $B_{p}$ shrinks to the 
point $p$ and the coefficients $g_{ij}$ tend to the constants 
$g_{ij}(p).$ On the other hand, the metrics $g_{\lambda}$ are defined 
on the intrinsic geodesic ball of radius $k$. Since $k$ is arbitrary, 
the metrics $\phi_{\lambda}^{*}g_{\lambda}$ converge smoothly to the 
limit flat metric $g_{0}$ on the tangent space $T_{p}(M),$ induced by 
the inner product $g_{p}$ on $T_{p}(M),$
\begin{equation} \label{e1.2}
(M, \phi_{\lambda}^{*}g_{\lambda}) \rightarrow (T_{p}M, g_{0}).
\end{equation}

 This process is called ``{\sf blowing up}'', since one restricts 
attention to smaller and smaller balls, and blows them up to a definite 
size. Note that the part of $M$ at any definite $g$-distance to $p$ 
escapes to infinity, and is not detected in the limit $g_{0}.$ Thus, it 
is important to attach base points to the blow-up construction; 
different base points may give rise to different limits, (although in 
this situation all pointed limits are isometric).

\medskip

 There is an analogous, although more subtle blowing up process for 
Lorentzian metrics due to Penrose, where the limits are non-flat plane 
gravitational waves, cf. [42].

\bigskip

 If $(M, g)$ is complete and non-compact, one can carry out a similar 
procedure with $\lambda  \rightarrow 0$, called ``{\sf blowing down}'', 
where geodesic balls, (about a given point), of large radius 
$B_{p}(k/\lambda )$ are rescaled down to unit size, i.e. size $k$. This 
is of importance in understanding the large scale or asymptotic 
behavior of the metric and will arise in later sections.

\medskip

 This discussion leads to the following definition for convergence of 
metrics.\index{convergence of metrics} Let $\Omega $ be a domain in 
${\mathbb R}^{n}$ and let $C^{k,\alpha}$ denote the usual H\"{o}lder space 
of $C^{k}$ functions on $\Omega$ with $\alpha$-H\"{o}lder continuous 
$k^{\rm th}$ partial derivatives. Similarly, let $L^{k,p}$ denote the Sobolev 
space of functions with $k$ weak derivatives in $L^{p}.$ Since one works only 
locally, we are only interested in the local spaces 
$C^{k,\alpha}_{loc}$ and $L^{k,p}_{loc}$ and corresponding local norms 
and topology.

\begin{definition} \label{d 1.1}
{\rm A sequence of metrics $g_{i}$ on $n$-manifolds $M_{i}$ is said to 
{\sf converge in the $L^{k,p}$ topology} to a limit metric $g$ on the 
$n$-manifold $M$ if there is a locally finite collection of charts 
$\{\phi_{k}\}$ covering $M$, and a sequence of diffeomorphisms $F_{i}: 
M \rightarrow  M_{i},$ such that 
\begin{equation} \label{e1.3}
(F_{i}^{*}g_{i})_{\alpha\beta} \rightarrow  g_{\alpha\beta}, 
\end{equation}
in the $L^{k,p}_{loc}$ topology. Here $(F_{i}^{*}g_{i})_{\alpha\beta}$ 
and $g_{\alpha\beta}$ are the local component functions of the metrics 
$F_{i}^{*}g_{i}$ and $g$ in the charts $\phi_{k}.$ }
\end{definition}

 The same definition holds for convergence in the $C^{k,\alpha}$ 
topology, as well as the weak $L^{k,p}$ topology. (Recall that a 
sequence of functions $f_{i}\in L^{p}(\Omega )$ converges weakly in 
$L^{p}$ to a limit $f \in  L^{p}(\Omega )$ iff $\int f_{i}g \rightarrow 
 \int fg,$ for all $g\in L^{q}(\Omega ),$ where $p^{-1}+q^{-1} =$ 1).

  It is easily seen that this definition of convergence is independent 
of the choice of charts $\{\phi_{k}\}$ covering $M$. The manifolds $M$ 
and $M_{i}$ are not required to be compact. When $M$ is non-compact, the 
convergence above is then uniform on compact subsets.

\bigskip

 In order to obtain local control on a metric, or sequence of metrics, 
one assumes curvature bounds. The theory described by Cheeger-Gromov 
requires a bound on the full Riemann curvature tensor 
\begin{equation} \label{e1.4}
|Riem| \leq  K, 
\end{equation}
for some $K <  \infty .$ Since the number of components of the Riemann 
curvature is much larger than that of the metric tensor itself, (in 
dimensions $\geq 4$), this corresponds to an overdetermined set of 
constraints on the metric and so is overly restrictive. It is much more 
natural to impose bounds on the Ricci curvature 
\begin{equation} \label{e1.5}
|Ric| \leq  k, 
\end{equation}
since the Ricci curvature is a symmetric bilinear form, just as the 
metric is. Of course, assuming bounds on Ricci is natural in general 
relativity, via the Einstein equations. Thus throughout the paper, we 
emphasize (1.5) over (1.4) whenever possible.

\bigskip

 The Cheeger-Gromov theory may be viewed as a vast generalization of 
the basic features of Teichm\"{u}ller theory to higher dimensions and 
variable curvature, (although it was not originally phrased in this 
way). Recall that Teichm\"{u}ller theory describes the moduli space 
$\mathcal{M}_{c}$ of constant curvature metrics on surfaces, cf. [46] 
and references therein. On closed surfaces, one has a 
{\it  basic trichotomy}  for the behavior of sequences of such metrics, 
normalized to unit area:

\smallskip

$\bullet$ {\it Compactness/Convergence}. A sequence $g_{i}\in 
\mathcal{M}_{c}$ has a subsequence converging smoothly, ($C^{\infty}$), 
to a limit metric $g\in \mathcal{M}_{c}.$ As in the definition above, 
the convergence is understood to be modulo diffeomorphisms. For 
instance this is always the case on $S^{2},$ since the moduli space 
$\mathcal{M}_{c}$ is a single point for $S^{2}$.

\smallskip

$\bullet$ {\it Collapse}. The sequence $g_{i}\in \mathcal{M}_{c}$ 
collapses everywhere, in that 
\begin{equation} \label{e1.6}
inj_{g_{i}}(x) \rightarrow 0,
\end{equation}
at every $x$, where $inj_{g_{i}}$ is the injectivity radius w.r.t. 
$g_{i}$. This collapse occurs only on the torus $T^{2}$ and such 
metrics become very long and very thin, (if the area is normalized to 1). 
There is no limit metric on $T^{2}.$ Instead, by choosing (arbitrary) base 
points $x_{i},$ one may consider based sequences $(T^{2}, g_{i}, x_{i})$, 
whose limits are then the ``collapsed'' space $({\mathbb R} , g_{\infty}, x_{\infty})$. 
Here ${\mathbb R}$ is the real line, and $g_{\infty}$ is any Riemannian metric on 
${\mathbb R}$; recall that all metrics on ${\mathbb R}$ are isometric. The 
convergence here is that of metric spaces, i.e. in the Gromov-Hausdorff 
topology, cf. [31], [43].

\smallskip

$\bullet$ {\it Cusp Formation}. This is a mixture of the two previous 
cases, and occurs only for hyperbolic metrics, i.e. on surfaces 
$\Sigma_{g}$ of genus $g \geq $ 2. In this case, there are based 
sequences $(\Sigma_{g}, g_{i}, x_{i})$ which converge to a limit 
$(\Sigma , g_{\infty}, x_{\infty})$ which is a complete non-compact 
hyperbolic surface of finite volume, hence with a finite number of cusp 
ends $S^{1}\times {\mathbb R}^{+}$. The convergence is smooth, and 
uniform on compact subsets. As one goes to infinity in any such cusp end 
$S^{1}\times {\mathbb R}^{+}$, the limit metric collapses in the sense 
that $inj_{g_{\infty}}(z_{k}) \rightarrow 0$, as $z_{k} \rightarrow 
\infty$. There are other based sequences $(\Sigma , g_{i}, y_{i})$ 
which collapse, i.e. (1.6) holds on domains of arbitrarily large but 
bounded diameter about $y_{i}$. As before, limits of such sequences are 
of the form $({\mathbb R} , g_{\infty}, y_{\infty}).$

\section{Convergence/Compactness.}
\setcounter{equation}{0}

 To prove the (pre)-compactness of a family of metrics, or the 
convergence of a sequence of metrics, the main point is to establish a 
lower bound on the radius of balls on which one has apriori control of 
the metric in a given topology, say $C^{k,\alpha}$ or $L^{k,p}.$ Given 
such uniform local control, it is then usually straightforward to 
obtain global control, via suitable global assumptions on the volume or 
diameter. (Alternately, one may work instead on domains of bounded 
diameter). 

 To obtain such local control, the first issue is to choose a good 
``gauge'', i.e. representation of the metric in local coordinates. For 
this, it is natural to look at coordinates built from the geometry of 
the metric itself. In the early stages of development of the theory, 
geodesic normal coordinates were used. Later, Gromov [31] used suitable 
distance coordinates. However, both these coordinate systems entail 
loss of derivatives - two in the former case, one in the latter. It is 
now well-known that Riemannian metrics have optimal regularity 
properties in harmonic coordinates, cf. [23]; this is due to the 
special form of the Ricci curvature in harmonic coordinates, known to 
relativists long ago.

 Given the choice of harmonic gauge, it is natural to associate a 
harmonic radius $r_{h}: M \rightarrow  {\mathbb R}^{+}$, which measures 
the size of balls on which one has harmonic coordinates in which the 
metric is well controlled. The precise definition, cf. [1], is as 
follows. \index{harmonic radius}
\begin{definition} \label{d2.1}
{\rm Fix a function topology, say $L^{k,p}$, and a constant $c_{o} >  
1$. Given $x\in (M, g)$, define the $L^{k,p}$ harmonic radius to be the 
largest radius $r_{h}(x) = r_{h}^{k,p}(x)$ such that on the geodesic ball 
$B_{x}(r_{h}(x))$ one has a harmonic coordinate chart $U = 
\{u_{\alpha}\}$ in which the metric $g = g_{\alpha\beta}$ is controlled 
in $L^{k,p}$ norm: thus,
\begin{equation} \label{e2.1}
c_{o}^{-1}\delta_{\alpha\beta} \leq  g_{\alpha\beta} \leq  
c_{o}\delta_{\alpha\beta}, 
\ \ {\rm (as \ bilinear \ forms)}, 
\end{equation}
\begin{equation} \label{e2.2}
[r_{h}(x)]^{kp-n}\int_{B_{x}(r_{h}(x))}|\partial^{k}g_{\alpha\beta}|^{p}
dV \leq  c_{o}-1. 
\end{equation}}
\end{definition}
Here, it always assumed that $kp > n = dim M$, so that $L^{k,p}$ embeds 
in $C^{0},$ via Sobolev embedding. The precise value of $c_{o}$ is 
usually unimportant, but is understood to be fixed once and for all. 
Both estimates in (2.1)-(2.2) are scale invariant, (when the harmonic 
coordinates are rescaled as in (1.1)), and hence the harmonic radius 
scales as a distance. 

\medskip

  Note that if $r_{h}(x)$ is large, then the metric is close to the 
flat metric on large balls about $x$, while if $r_{h}(x)$ is small, 
then the derivatives of $g_{\alpha \beta}$ up to order $k$ are large in 
$L^{p}$ on small balls about $x$. Thus, the harmonic radius serves as a 
measure of the degree of concentration of $g_{\alpha \beta}$ in the 
$L^{k,p}$ norm.

 It is important to observe that the harmonic radius is continuous with 
respect to the (strong) $L^{k,p}$ topology on the space of metrics, cf. 
[1], [3]. In general, it is not continuous in the weak $L^{k,p}$ 
topology. 

 One may define such harmonic radii w.r.t. other topologies, for 
instance $C^{k,\alpha}$ in a completely analogous way; these have the 
same properties.

\medskip

 Suppose $g_{k}$ is a sequence of metrics on a manifold $M$, (possibly 
open), with a uniform lower bound on $r_{h}.$ On each ball, one then 
has $L^{k,p}$ control of the metric components. The well-known 
Banach-Alaoglu theorem, (bounded sequences are weakly compact in reflexive 
Banach spaces), then implies that the metrics on the ball have a weakly 
convergent subsequence in $L^{k,p},$ so one obtains a limit metric on 
each ball. Using elliptic regularity associated with harmonic functions, 
it is straightforward to verify that the overlaps of these 
charts are in $L^{k+1,p}$, and so one has a limit $L^{k,p}$ metric on 
$M$. The convergence to limit is in the weak $L^{k,p}$ topology and 
uniform on compact subsets. Strictly speaking, one also has to prove 
that the harmonic coordinate charts for $g_{k}$ also converge, or more 
precisely may be replaced by a fixed coordinate chart, but this also is 
not difficult, cf. [1], [3] for details.

 The same type of arguments hold w.r.t. the $C^{k,\alpha}$ topology, 
via the Arzela-Ascoli theorem; here weak $L^{k,p}$ convergence is 
replaced by convergence in the $C^{k,\alpha'}$ topology, for $\alpha'  
<  \alpha$.

\medskip

 Thus, the main issue in obtaining a convergence result is to obtain a 
lower bound on a suitable harmonic radius $r_{h}$ under geometric 
bounds. The following result from [1] is one typical example.

\begin{theorem} \label{t 2.2} {\bf (Convergence I).}
Let $M$ be a closed $n$-manifold and let $\mathcal{M} (\lambda , i_{o}, 
D)$ be the space of Riemannian metrics such that
\begin{equation} \label{e2.3}
|Ric| \leq  k, \ inj \geq  i_{o}, \ diam \leq  D. 
\end{equation}
Then $\mathcal{M} (\lambda , i_{o},$ D) is precompact in the 
$C^{1,\alpha}$ and weak $L^{2,p}$ topologies, for any $\alpha  < $ 1 
and $p <  \infty$.
\end{theorem}
 Thus, for any sequence, there is a subsequence which converges, in 
these topologies, to a limit $C^{1,\alpha}\cap L^{2,p}$ metric 
$g_{\infty}$ on $M$.

\medskip
\noindent
{\bf Sketch of Proof}: As discussed above, it suffices to prove a 
uniform lower bound on the $L^{2,p}$ harmonic radius $r_{h} = 
r_{h}^{2,p},$ i.e.
\begin{equation} \label{e2.4}
r_{h}(x) \geq  r_{o} = r_{o}(k, i_{o}, D), 
\end{equation}
under the bounds (2.3). 

 Overall, the proof of (2.4) is by contradiction. Thus, if (2.4) is 
false, there is a sequence of metrics $g_{i}$ on $M$, satisfying the 
bounds (2.3), but for which $r_{h}(x_{i}) \rightarrow $ 0, for some 
points $x_{i}\in M$. Without loss of generality, (since $M$ is closed), 
assume that the base points $x_{i}$ realize the minimal value of 
$r_{h}$ on $(M, g_{i}).$ Then rescale the metrics $g_{i}$ by this 
minimal harmonic radius, i.e. set
\begin{equation} \label{e2.5}
\bar g_{i} = r_{h}(x_{i})^{-2}\cdot  g_{i}. 
\end{equation}
If $\bar r_{h}$ denotes the harmonic radius w.r.t. $\bar g,$ by scaling 
properties one has
\begin{equation} \label{e2.6}
\bar r_{h}(x_{i}) = 1, \ \ {\rm and} \ \ \bar r_{h}(y_{i}) \geq  1, 
\end{equation}
for all $y_{i}\in (M, \bar g_{i}).$ By the remarks preceeding the 
proof, the pointed Riemannian manifolds $(M, \bar g_{i}, x_{i})$ have a 
subsequence converging in the {\it  weak}  $L^{2,p}$ topology to a 
limit $L^{2,p}$ Riemannian manifold $(N, \bar g_{\infty}, x_{\infty}).$ 
(Again, this convergence is understood to be modulo diffeomorphisms, as 
in Definition 1.1). Of course $diam_{\bar g_{i}}M \rightarrow  \infty 
,$ so that the complete open manifold $N$ is distinct from the original 
compact manifold $M$. The convergence is uniform on compact subsets.

 So far, nothing essential has been done - the construction above more 
or less amounts to just renormalizations. There are two basic 
ingredients in obtaining further control however, one geometric and one 
analytic.

 We begin with the geometric argument. The limit space $(N, \bar 
g_{\infty})$ is Ricci-flat, since the bound (2.3) on the Ricci 
curvature of $g_{i}$ becomes in the scale $\bar g_{i},$
\begin{equation} \label{e2.7}
|Ric_{\bar g_{i}}| \leq  k\cdot  r_{h}(x_{i}) \rightarrow  0, \ \ {\rm 
as} \ \ i \rightarrow  \infty . 
\end{equation}
 Actually, it is Ricci-flat in a weak sense, since the convergence is 
only in weak $L^{2,p}.$ However, it is easy to see, (cf. also below), 
that weak $L^{2,p}$ solutions of the (Riemannian) Einstein equations 
are real-analytic, and so the limit is in fact a smooth Ricci-flat 
metric.

 Next, by (2.3), the injectivity radius of $\bar g_{i}$ satisfies
\begin{equation} \label{e2.8}
inj_{\bar g_{i}} \geq  i_{o}\cdot  r_{h}(x_{i})^{-1} \rightarrow  
\infty , \ \ {\rm as} \ \ i \rightarrow  \infty , 
\end{equation}
so that, roughly speaking, the limit $(N, \bar g_{\infty})$ has 
infinite injectivity radius at every point. More importantly, the bound 
(2.8) implies that $(M, \bar g_{i})$ contains arbitrarily long, 
(depending on $i$), minimizing geodesics in any given direction through 
the center point $x_{i}.$ It follows that the limit $(N, \bar 
g_{\infty})$ has infinitely long minimizing geodesics in every 
direction through the base point $x_{\infty}.$ This means that $(N, 
\bar g_{\infty})$ contains a line in every direction through 
$x_{\infty}.$ 

 Now the well-known Cheeger-Gromoll splitting theorem [17] \index{splitting 
theorem} states that a complete manifold with non-negative Ricci curvature 
splits isometrically along any line. It follows that $(N, \bar g_{\infty})$ 
splits isometrically in every direction through $x_{\infty},$ and hence 
$(N, \bar g_{\infty}) = ({\mathbb R}^{n}, g_{0}),$ where $g_{0}$ is the 
flat metric on ${\mathbb R}^{n}.$

 Now of course $({\mathbb R}^{n}, g_{0})$ has infinite harmonic radius. 
If the convergence of $(N, \bar g_{i})$ to the limit $({\mathbb R}^{n}, 
g_{0})$ can be shown to be in the {\sf strong} $L^{2,p}$ topology, then 
the continuity of $r_{h}$ in this topology immediately gives a 
contradiction, since by (2.6), the limit $(N, \bar g_{\infty})$ has 
$r_{h}(x_{\infty}) = 1$.

\medskip

 The second or analytic part of the argument is to prove strong 
$L^{2,p}$ convergence to the limit. The idea here is to use elliptic 
regularity to bootstrap or improve the smoothness of the convergence.

 In harmonic coordinates, the Ricci curvature of a metric $g$ has the 
following especially simple form:
\begin{equation} \label{e2.9}
-\frac{1}{2}\Delta g_{\alpha\beta} + Q_{\alpha\beta}(g, \partial g)= 
Ric_{\alpha\beta}, 
\end{equation}
where $\Delta  = g^{\alpha\beta}\partial_{\alpha}\partial_{\beta}$ is 
the Laplacian w.r.t. the metric $g$ and $Q$ is quadratic in $g$, its 
inverse, and $\partial g.$ In particular, if $r_{h}(x) =$ 1 and 
$r_{h}(y) \geq  r_{o} > $ 0, for all $y\in\partial B_{x}(1),$ then one 
has a uniform $L^{1,p}$ bound on $Q$ and uniform $L^{2,p}$ bounds on 
the coefficients for the Laplacian within $B_{x}(1+\frac{1}{2}r_{o}).$ 

 If now $Ric$ is uniformly bounded in $L^{\infty},$ then standard 
elliptic regularity applied to (2.9) implies that $g_{\alpha\beta}$ is 
uniformly controlled in $L^{2,q},$ for any $q <  \infty$, (in 
particular for $q > p$). More importantly, if $g_{i}$ is a sequence of 
metrics for which $(Ric_{g_{i}})_{\alpha\beta}$ converges strongly in 
$L^{p}$ to a limit $(Ric_{g_{\infty}})_{\alpha\beta},$ then elliptic 
regularity again implies that the metrics $(g_{i})_{\alpha\beta}$ 
converge strongly in $L^{2,p}$ to the limit 
$(g_{\infty})_{\alpha\beta}.$ For the metrics $\bar g_{i}$, (2.7) 
implies that $Ric \rightarrow 0$ in $L^{\infty}$, and so $Ric 
\rightarrow 0$ strongly in $L^{q}$, for any $q < \infty$.

 These remarks essentially prove that the $L^{2,p}$ harmonic radius is 
continuous w.r.t. the strong $L^{2,p}$ topology. Further, when applied 
to the sequence $\bar g_{i}$ and using (2.6), they imply that the 
metrics $\bar g_{i}$ converge strongly in $L^{2,p}$ to the limit $\bar 
g_{\infty}.$ This completes the proof.

\bigskip

 It is easy to see from the proof that the lower bound on the 
injectivity radius in (2.3) can be considerably weakened. For instance, 
define the 1-cross $Cro_{1}(x)$ of $(M, g)$ at $x$ to be the length of 
the longest minimizing geodesic in $(M, g)$ with center point $x$ and 
set 
$$Cro_{1}(M,g) = \inf_{x}Cro_{1}(x).$$
We introduce this notion partly because it has a natural analogue in 
Lorentzian geometry, when a minimizing geodesic is replaced by a 
maximizing time-like geodesic, cf. \S 5. Then one has the following 
result on 4-manifolds, cf. [4].

\begin{theorem} \label{t 2.3} {\bf (Convergence II).}
Let $M$ be a 4-manifold. Then the conclusions of Theorem 2.2 hold under 
the bounds
\begin{equation} \label{e2.10}
|Ric| \leq  k, \ Cro_{1} \geq  c_{o}, \ vol \geq  v_{o}, \ diam \leq  
D. 
\end{equation}
\end{theorem}

 The proof is the same as that of Theorem 2.2. The lower bound on 
$Cro_{1}$ implies that on the blow-up limit $(N, \bar g_{\infty}, 
x_{\infty})$ above, one has {\it a} line. Hence, the splitting theorem 
implies that $N = N' \times {\mathbb R}$. It follows that $N'$ is 
Ricci-flat and hence, since $dim N' = 3$, $N'$ is flat. Using the 
volume bound in (2.10), it follows that $(N, \bar g_{\infty}) = 
({\mathbb R}^{4}, g_{0})$, cf. (2.12)-(2.13) below. (The volume bound 
rules out the possibility that $N'$ is a non-trivial flat manifold of 
the form ${\mathbb R}^{3}/\Gamma$). This gives the same contradiction 
as before. 
 
\medskip

 Of course, in dimension 3 any Ricci-flat manifold is necessarily flat, 
and so the same proof shows that one has $C^{1,\alpha}$ and $L^{2,p}$ 
precompactness within the class of metrics on 3-manifolds satisfying
\begin{equation} \label{e2.11}
|Ric| \leq  k, \ vol \geq  v_{o}, \ diam \leq  D. 
\end{equation}
Thus, no assumptions on $inj$ or $Cro_{1}$ are needed in dimension 3.

\begin{remark} \label{r 2.4.} {\bf (i).} 
{\rm Although (2.4) gives the existence of a lower bound on $r_{h}$ in 
terms of the bounds $k$, $i_{o}$ and $D$, currently there is no proof 
of an effective or computable bound. Equivalently, there is no direct 
proof of Theorem 2.2, which does not involve a passage to limits and 
invoking a contradiction. This is closely related to the fact there is 
currently no {\it quantitative}  or {\it  finite}  version of the 
Cheeger-Gromoll splitting theorem, where one can deduce definite bounds 
on the metric in the presence of (a collection of) minimizing geodesics 
of a finite but definite length.

  If however the bound on $|Ric|$ in (2.3) is strengthened to a bound 
on $|Riem|$, as in (1.4), then it is not difficult to obtain an 
effective or computable lower bound on $r_{h}$, cf. [36].}

 {\bf (ii).} {\rm The proof above can be easily adapted to give a 
similar result if the $L^{\infty}$ bound on $Ric$ is replaced by an 
$L^{q}$ bound, for some $q >  n/2;$ one then obtains convergence in 
weak $L^{2,q}.$

 In the opposite direction, the convergence can be improved if one has 
bounds on the derivatives of the Ricci curvature. This will be the case 
if $Ric$ satisfies an elliptic system of PDE, for instance the Einstein 
equations. In this case, one obtains $C^{\infty}$ convergence to the 
limit. }

  {\bf (iii).} {\rm The assumption that $M$ is closed in Theorem 2.2 is 
merely for convenience, and an analogous result holds for open 
manifolds, away from the boundary. }

\end{remark}

 The bounds on injectivity radius in (2.3), or even the 1-cross in 
(2.10), are rather strong and one would like to replace them with 
just a lower volume bound, as in (2.11).

 An elementary but important result, the volume comparison theorem of 
Bishop-Gromov [31], [43], states that if $Ric \geq (n-1)k$, for some $k$, on 
$(M, g)$, $n = dim M$, then the ratio
\begin{equation} \label{e2.12}
\frac{volB_{x}(r)}{volB_{k}(r)} 
\end{equation}
is monotone non-increasing in $r$; here $volB_{k}(r)$ is the volume of 
the geodesic $r$-ball in the $n$-dimensional space form of constant 
curvature $k$. In particular, if the bounds (2.11) hold, in dimension 
$n$, then (2.12) gives a lower bound on the volumes of balls on {\it 
all} scales:
\begin{equation} \label{e2.13}
volB_{x}(r) \geq  \frac{volM}{volB_{k}(D)}\cdot  volB_{k}(r). 
\end{equation}
Note that the estimate (2.13) also implies that, for any fixed $r > 0$, 
if $volB_{x}(r) \geq v_{0} > 0$, then  $volB_{y}(r) \geq v_{1} > 0$, 
where $v_{1}$ depends only on $v_{0}$ and $dist_{g}(x, y)$. Thus, the 
ratio of the volumes of unit balls cannot become arbitrarily large or 
small on domains of bounded diameter.

 Now a classical result of Cheeger [16] implies that if (2.11) is 
strengthened to
\begin{equation} \label{e2.14}
K_{P} \geq  - K, \ vol \geq  v_{o}, \ diam \leq  D, 
\end{equation}
where $K_{P}$ is the sectional curvature of any plane $P$ in the 
tangent bundle $TM$, then one has a lower bound on the injectivity 
radius, $inj_{g}(M) \geq  i_{o}(K, v_{o}, D)$. However, it was observed 
in [2] that this estimate fails under the bounds (2.11). It is worthwhile 
to exhibit a simple concrete example illustrating this.

\begin{example} \label{ex 2.5}
{\rm Let $g_{\lambda}$ be the family of Eguchi-Hanson metrics on the 
tangent bundle $TS^{2}$ of $S^{2}.$ The metrics $g_{\lambda}$ are given 
explicitly by
\begin{equation} \label{e2.15}
g_{\lambda} = [1-(\frac{\lambda}{r})^{4}]^{-1}dr^{2} + 
r^{2}[1-(\frac{\lambda}{r})^{4}]\theta_{1}^{2} + 
r^{2}(\theta_{2}^{2}+\theta_{3}^{2}). 
\end{equation}
Here $\theta_{1}, \theta_{2}, \theta_{3}$ are the standard 
left-invariant coframing of $SO(3) = {\mathbb R}{\mathbb P}^{3}$, (the 
sphere bundles in $TS^{2})$ and $r \geq  \lambda .$ The locus 
$r=\lambda $ is the image of the 0-section and is a totally geodesic 
round $S^{2}(\lambda )$ of radius $\lambda .$ 

 The metrics $g_{\lambda}$ are Ricci-flat, and are all homothetic, i.e. 
are rescalings (via diffeomorphisms) of a fixed metric; in fact,
\begin{equation} \label{e2.16}
g_{\lambda} = \lambda^{2}\cdot \psi_{\lambda}^{*}(g_{1}), 
\end{equation}
where $\psi_{\lambda}(r) = \lambda r,$ and $\psi_{\lambda}$ acts 
trivially on the $SO(3)$ factor. As $\lambda  \rightarrow $ 0, i.e. as 
one blows down the metrics, $g_{\lambda}$ converges to the metric 
$g_{0},$ the flat metric on the cone $C({\mathbb R}{\mathbb P}^{3}).$ 
The convergence is smooth in the region $r \geq  r_{o},$ for any fixed 
$r_{o} > 0$, but is not smooth at $r =$ 0. Since $S^{2}(\lambda )$ is 
totally geodesic, the injectivity radius at any point of $S^{2}(\lambda 
)$ is $2\pi\lambda ,$ which tends to 0. On the other hand, the volumes 
of unit balls, or balls of any definite radius, remain uniformly 
bounded below. }
\end{example}

 One sees here that the metrics $(TS^{2}, g_{\lambda})$ converge as 
$\lambda  \rightarrow $ 0 to a limit metric on a singular space 
$C({\mathbb R}{\mathbb P}^{3}).$ The limit is an orbifold ${\mathbb 
R}^{4}/{\mathbb Z}_{2},$ where ${\mathbb Z}_{2}$  acts by reflection in 
the origin.

\medskip

 The Eguchi-Hanson metric is the first and simplest example of a large 
class of Ricci-flat ALE (asymptotically locally Euclidean) spaces, 
whose metrics are asymptotic to cones $C(S^{3}/\Gamma )$, $\Gamma  
\subset SO(4)$, on spherical space forms. This is the family of ALE 
gravitational instantons, studied in detail by Gibbons and Hawking, cf. 
[30] and references therein, in connection with Hawking's Euclidean 
quantum gravity program.

\medskip

 It is straightforward to modify the construction in Example 2.5 to 
obtain orbifold degenerations on compact 4-manifolds satisfying the 
bounds (2.11). Thus, one does not have $C^{1,\alpha}$ or even $C^{0}$ 
(pre)-compactness of the space of metrics on $M$ under the bounds 
(2.11). Singularities can form in passing to limits, although the 
singularities are of a relatively simple kind. The next result from [1] 
shows that this is the only kind of possible degeneration or 
singularity formation.

\begin{theorem} \label{t 2.6} {\bf (Convergence III).}
Let $\{g_{i}\}$ be a sequence of metrics on a 4-manifold, satisfying 
the bounds 
\begin{equation} \label{e2.17}
|Ric| \leq  k, \ vol \geq  v_{o}, \ diam \leq  D. 
\end{equation}
Then a subsequence converges, (in the Gromov-Hausdorff topology), to an 
orbifold $(V, g)$, with a finite number of singular points $\{q_{j}\}.$ 
Each singular point $q$ has a neighborhood homeomorphic to a cone 
$C(S^{3}/\Gamma ),$ for $\Gamma $ a finite subgroup of $SO(4)$. 

 The metric $g$ is $C^{1,\alpha}$ or $L^{2,p}$ on the regular set 
$$V_{0} = V \setminus \cup\{q_{j}\},$$
and extends in a local uniformization of a singular point to a $C^{0}$ 
Riemannian metric. Further, there are embeddings 
$$F_{i}: V_{0} \rightarrow  M$$
such that $F_{i}^{*}(g_{i})$ converges in the $C^{1,\alpha}$ topology 
to the metric $g$.
\end{theorem}

 Here, convergence in the Gromov-Hausforff topology means convergence 
as metric spaces, cf. [31], [43]. We mention only a few important issues in 
the proof of Theorem 2.6. First, the Chern-Gauss-Bonnet formula implies 
that for metrics with bounded Ricci curvature and volume on 
$4$-manifolds, one has an apriori bound on the $L^{2}$ norm of the full 
curvature tensor:
$${\textstyle \frac{1}{8\pi^{2}}}\int_{M}|R|^{2}dV \leq \chi (M) + C(k, 
V_{o}), $$
where $C(k, V_{o})$ is a constant depending only on $k$ from (2.17) and 
an upper bound $V_{o}$ on $vol_{g}M$: $\chi (M)$ is the Euler 
characteristic of $M$. Second, with each singular point $q \in  V$, 
there is a associated a sequence of rescalings $\bar g_{i} = 
\lambda_{i}^{2}g_{i}, \lambda_{i} \rightarrow  \infty ,$ and base 
points $x_{i}\in M$, $x_{i} \rightarrow q$, such that a subsequence of 
$(M, \bar g_{i}, x_{i})$ converges in $C^{1,\alpha} \cap L^{2,p}$ to a 
non-trivial Ricci-flat ALE space $(N, \bar g_{\infty})$ as above. It is 
not difficult to see that any such ALE space has a definite amount of 
curvature in $L^{2}.$ This implies basically that there are only a 
finite number of such singular points. Further, the ALE spaces $N$ are 
embedded in $M$, in a topologically essential way.

\section{Collapse/Formation of Cusps.}
\setcounter{equation}{0}

 In this section, we consider what happens when 
$$vol \rightarrow  0 \ \ {\rm or} \ \ diam \rightarrow  \infty$$
in the bounds (2.11). This involves the notion of Cheeger-Gromov 
collapse, or collapse with bounded curvature. \index{collapse of metrics} 

 For simplicity, we restrict the discussion to dimension 3. While there 
is a corresponding theory in higher dimensions, cf. [18], there are 
special and advantageous features that hold only in dimension 3 in general. 
Further, the relations with general relativity are most direct in 
dimension 3, in that the discussion can be applied to the behavior of 
space-like hypersurfaces in a given space-time.

 The simplest non-trivial example of collapse is the Berger collapse of 
the 3-sphere along $S^{1}$ fibers of the Hopf fibration. Thus, consider 
the family of metrics on $S^{3}$ given by
\begin{equation} \label{e3.1}
g_{\lambda} = \lambda^{2}\theta_{1}^{2} + 
(\theta_{2}^{2}+\theta_{3}^{2}), 
\end{equation}
where $\theta_{1}, \theta_{2}, \theta_{3}$ are the standard 
left-invariant coframing of $S^{3}.$ The metrics $g_{\lambda}$ have an 
isometric $S^{1}$ action, with Killing field $K$ dual to $\theta_{1},$ 
with length of the $S^{1}$ orbits given by $2\pi\lambda .$ Thus, in 
letting $\lambda  \rightarrow $ 0, one is blowing down the metric in 
{\it one}  direction. (This is exactly what occurs on approach to the 
horizon of the Taub-NUT metric, cf. [35]). A simple calculation shows 
that the curvature of $g_{\lambda}$ remains uniformly bounded as 
$\lambda  \rightarrow 0$. Clearly $vol_{g_{\lambda}}S^{3} \sim \lambda 
\rightarrow 0$. The metrics $g_{\lambda}$ collapse $S^{3}$ to a limit 
space, in this case $S^{2}.$

 This same procedure may be carried out, with the same results, on any 
3-manifold (or $n$-manifold) which has a free or locally free isometric 
$S^{1}$ action; locally free means that the isotropy group of any orbit 
is a finite subgroup of $S^{1},$ i.e. there are no fixed points of the 
action. Similarly, one may collapse along the orbits, as in (3.1), of a 
locally free $T^{k}$-action, where $T^{k}$ is the $k$-torus. 
Remarkably, Gromov [32] showed that more generally one may collapse 
along the orbits of an isometric nilpotent group action, and 
furthermore, such groups are {\it only} groups which allow such a 
collapse with bounded curvature. Thus for instance collapsing along the 
orbits of an isometric $G$-action, where $G$ is semi-simple and 
non-abelian, increases the curvature without bound.

\medskip

 A 3-manifold which admits a locally free $S^{1}$ action is called a 
{\sf Seifert fibered space}. Such a space admits a fibration over a 
surface $V$, with $S^{1}$ fibers. Where the action is free, this 
fibration is a circle bundle. There may exist an isolated collection of 
non-free orbits, corresponding to isolated points in $V$. 
Topologically, a neighborhood of such an orbit is of the form $D^{2} 
\times S^{1}$, where the $S^{1}$ acts by rotation on the $S^{1}$ factor 
and by rotation through a rational angle about $\{0\}$ in $D^{2}$.

  The collection of Seifert fibered spaces falls naturally into 6 
classes, according to the topology of the base surface $V$, i.e. $V = 
S^{2}$, $T^{2}$, or $\Sigma_{g}$, $g \geq 2$, and according to whether 
the $S^{1}$ bundle is trivial or not trivial. These account for 6 of 
the 8 possible geometries of 3-manifolds in the sense of Thurston [46]. 
These geometries are: $S^{2} \times {\mathbb R}$, ${\mathbb R}^{3}$, 
${\mathbb H}^{2} \times {\mathbb R}$, $S^{3}$, $Nil$, and 
$\widetilde{SL(2, {\mathbb R})}$, respectively. The two remaining 
geometries are $Sol$, corresponding to non-trivial torus bundles over 
$S^{1}$, and the hyperbolic geometry ${\mathbb H}^{3}$.

\medskip

 Now suppose $N$ is a compact Seifert fibered space with boundary. The 
boundary is a finite collection of tori, on which one has a free 
$S^{1}$ action. In a neighborhood of the boundary, this $S^{1}$ action 
then embeds in the standard free $T^{2}$ action on $T^{2}\times I.$ 
Given a collection of such spaces $N_{i},$ one may then glue the toral 
boundaries together by automorphisms of the torus, i.e. by elements of 
$SL(2, {\mathbb Z})$. For example, the glueing may interchange the 
fiber and base circles.

\begin{definition} \label{d 3.1}
A {\sf graph manifold} \index{graph manifold} $G$ is a 3-manifold obtained 
by glueing Seifert fibered spaces by toral automorphisms of the boundary tori.
\end{definition}

 Thus, a graph manifold has a decomposition into two types of regions, 
\begin{equation} \label{e3.2}
G = S \cup L.
\end{equation}
Each component of $S$ is a Seifert fibered space, while each component 
of $L$ is $T^{2}\times I,$ and glues together different boundary 
components of elements in $S$. The exceptional case of glueing two 
copies of $T^{2}\times I$ by toral automorphisms of the boundary is 
also allowed; this defines the class of $Sol$ manifolds, up to finite 
covers. The Seifert fibered components have a locally free $S^{1}$ 
action, the $T^{2}\times I$ components have a free $T^{2}$ action; in 
general, these group actions do not extend to actions on topologically 
larger domains. 

  Graph manifolds are an especially simple class of 3-manifolds; they were 
introduced, and their structure was completely classified, by Waldhausen 
[48]. The terminology comes from the fact that one may associate a graph to 
$G$, by assigning a vertex to each component of $S$, and an edge to 
each component of $L$ which connects a pair of components in $S$.

\medskip

 It is not difficult to generalize the construction above to show that 
any closed graph manifold $G$ admits a sequence of metrics $g_{i}$ 
which collapse with uniformly bounded curvature, i.e.
\begin{equation} \label{e3.3}
|Ric_{g_{i}}| \leq  k, \ \ vol_{g_{i}}G \rightarrow  0. 
\end{equation}
The metrics $g_{i}$ collapse the Seifert fibered pieces along the 
$S^{1}$ orbits, while collapsing the toral regions $T^{2} \times I$ 
along the tori. Thus the collapse is rank 1 along $S$, while rank 2 
along $L$. (Of course a bound on the full curvature is the same as a 
bound on the Ricci curvature in dimension 3).

 If the graph manifold is Seifert fibered, then the collapse (3.3) may 
be carried out with bounded diameter,
\begin{equation} \label{e3.4}
diam_{g_{i}}S \leq D, \ {\rm for \ some} \ D < \infty.
\end{equation}
In fact, if $S$ is a $Nil$-manifold, then the collapse may be carried 
out so that $diam_{g_{i}}S \rightarrow 0$, cf. [32].

 On the other hand, suppose $G$ is non-trivial in that it has both $S$ 
and $L$ components. If $N$ denotes any $S$ or $L$ component, then it follows 
from work of Fukaya [27] that 
\begin{equation} \label{e3.5}
diam_{g_{i}}N \rightarrow  \infty  
\end{equation}
under the bounds (3.3). This phenomenon can be viewed as a refinement 
of the remark following (2.13), in that one has uniform control on the 
relative size of the injectivity radius on domains of bounded diameter, 
(cf. also [31]). In particular, the transition from 
Seifert fibered domains to toral domains takes longer and longer 
distance the more collapsed the metrics are. One obtains different 
collapsed ``limits'' depending on choice of base point. This ``pure'' 
behavior on regions of bounded diameter is special to dimension 3, cf. 
[5] for further details. 

 The discussion above shows that one may collapse graph manifolds with 
bounded curvature. The Cheeger-Gromov theory, [18], see also [33], implies 
that the converse also holds.

\begin{theorem} \label{t 3.2} {\bf (Collapse).}
If $M$ is a closed 3-manifold which collapses with bounded curvature, 
i.e. there is a sequence of metrics such that (3.3) holds, then $M$ is 
a graph manifold.
\end{theorem}

 In fact, this result holds if $M$ admits a sufficiently collapsed 
metric, i.e. $|Ric_{g}| \leq  k$ and $vol_{g}M \leq  \varepsilon_{o}$, 
for some $\varepsilon = \varepsilon_{o}(k)$ sufficiently small. Note of 
course that a collapsing sequence of metrics $g_{i}$ is not {\it  
necessarily} invariant under the $S^{1}$ or $T^{2}$ actions associated with 
the graph manifold structure; these local group actions are smooth actions, 
but need not be isometric w.r.t. a highly collapsed metric. 

\medskip

  In a certain sense, the vast majority of 3-manifolds are not graph 
manifolds, and so Theorem 3.2 gives strong topological restrictions on 
the existence of sufficiently collapsed metrics.

\medskip
\noindent
{\bf Idea of proof}: First, it is easy to see from elementary comparison 
geometry, cf. [43], that
$$vol_{g_{i}}B_{x}(1) \rightarrow 0 \Rightarrow inj_{g_{i}}(x) 
\rightarrow 0.$$
At any $x$, rescale the metrics $g_{i}$ to make $inj(x) = 1$, i.e. set
$$\bar g_{i} = [inj_{g_{i}}(x)]^{-2} \cdot g_{i}.$$
Now the bound (3.3) gives $|Ric_{\bar g_{i}}| \sim 0$. Thus, the 
metrics $\bar g_{i}$ are close to flat metrics on ${\mathbb 
R}^{3}/\Gamma$, where $\Gamma$ is a non-trivial discrete group of 
Euclidean isometries, (by Theorem 2.2 for instance). Thus, essentially, 
${\mathbb R}^{3}/\Gamma = {\mathbb R}^{2} \times S^{1}$, or ${\mathbb 
R} \times S^{1} \times S^{1}$. It follows that the local geometry, i.e. 
the geometry on the scale of the injectivity radius, is modeled by {\it 
non-trivial, flat} 3-manifolds. One then shows that these local 
structures for the geometry and topology can be glued together 
consistently to give a global graph manifold structure.

\medskip

 If $S$ is a closed Seifert fibered space, the orbits of the $S^{1}$ action 
always inject in $\pi_{1}(S)$, i.e. 
$$\pi_{1}(S^{1}) \hookrightarrow \pi_{1}(S),$$
unless $S = S^{3}/\Gamma $. In case $S$ has non-empty toral boundary 
components, the tori in $\partial S$ always inject in $\pi_{1}(S)$ except in 
the single case of $S = D^{2}\times S^{1},$ cf. [41]. Thus, if a graph 
manifold $G$ is not a spherical space form, or does not have a solid torus 
component in its Seifert fibered decomposition (3.2), then the fibers of the 
decomposition, namely circles and tori, always inject in $\pi_{1}$:
\begin{equation} \label{e3.6}
\pi_{1}(fiber) \hookrightarrow \pi_{1}(G).
\end{equation}
 Hence, in this situation, one can pass to covering spaces to {\it unwrap} 
any collapse. If $g_{i}$ is a collapsing sequence of 
metrics, by passing to larger and larger covering spaces, based 
sequences will always have convergent subsequences (in domains of 
arbitrary but bounded diameter). In addition, the isometric covering 
transformations on the covers have displacement functions converging 
uniformly to 0 on compact subsets. Hence, all such limits have a free 
isometric $S^{1}$ or $T^{2}$ action, depending on whether the collapse 
is rank 1 or 2 on the domains. This means that the limits have an {\sf 
extra symmetry} not necessarily present on the initial collapsing 
sequence. Again, this feature of being able to unwrap collapse by 
passing to covering spaces is special to dimension 3, cf. [5] for 
further discussion and applications.

\bigskip

 Finally, we discuss the third possibility, the formation of cusps.
\index{formation of cusps} 
This case, although the most general, corresponds to a mixture of the 
two previous cases convergence/collapse, and so no essentially new 
phenomenon occurs. To start, given a complete Riemannian manifold $(M, 
g)$, choose $\varepsilon  > $ 0 small, and let
\begin{equation} \label{e3.7}
M^{\varepsilon} = \{x\in M: volB_{x}(1) \geq  \varepsilon\}, \ 
M_{\varepsilon} = \{x\in M: volB_{x}(1) \leq  \varepsilon\}. 
\end{equation}
$M^{\varepsilon}$ is called the $\varepsilon$-thick part of $(M, g)$, 
while $M_{\varepsilon}$ is the $\varepsilon$-thin part.\index{thick-thin 
decomposition}

 Now suppose $g_{i}$ is a sequence of complete Riemannian metrics on 
the manifold $M$. 

\noindent
$\bullet$ If $x_{i} \in M^{\varepsilon}$, for some fixed $\varepsilon > 
0$, then one has convergence, (in subsequences), in domains of 
arbitrary but bounded diameter about $\{x_{i}\}$, see the discussion 
concerning (2.13). Essentially, the bounds (2.11) hold on such domains in 
this case.

\noindent
$\bullet$ If $y_{i} \in M_{\varepsilon_{o}}$, for $\varepsilon_{o}$ 
sufficiently small, then domains of bounded, depending on 
$\varepsilon_{o}$, diameter about $\{y_{i}\}$ are graph manifolds, in 
fact Seifert fibered spaces.

\noindent
$\bullet$ If $z_{i} \in M_{\varepsilon_{i}}$, $\varepsilon_{i} 
\rightarrow 0$, then domains of arbitrary but bounded diameter about 
$\{z_{i}\}$ are collapsing. 

\medskip

  If $(M_{\varepsilon}, g_{i}) = \emptyset$, for some fixed 
$\varepsilon > 0$, then one is in the convergence situation. If 
$(M^{\varepsilon}, g_{i}) = \emptyset$, for all $\varepsilon > 0$ 
sufficiently small, depending on $i$, then one is in the collapsing 
situation. The only remaining possibility is that, for any fixed small 
$\varepsilon > 0$,
\begin{equation} \label{e3.8}
(M^{\varepsilon}, g_{i}) \neq \emptyset, \ {\rm and} \ 
(M_{\varepsilon}, g_{i}) \neq \emptyset .
\end{equation} 
This is equivalent to the existence of base points $x_{i}$, $y_{i}$, 
such that,
\begin{equation} \label{e3.9}
vol B_{x_{i}}(1) \geq  \varepsilon_{1}, \ \ vol B_{y_{i}}(1) 
\rightarrow  0, 
\end{equation}
for some $\varepsilon_{1} > 0$. Observe that the volume comparison 
theorem (2.13) implies that $dist_{g_{i}}(x_{i}, y_{i}) \rightarrow  
\infty $ as $i \rightarrow  \infty ,$ so that these different behaviors 
become further and further distant as $i \rightarrow  \infty .$

 This leads to the following result, cf. [5], [18] for further details.

\begin{theorem} \label{t 3.3} {\bf (Cusp Formation).}
Let $M$ be a 3-manifold and $g_{i}$ a sequence of unit volume metrics 
on $M$ with uniformly bounded curvature, and satisfying (3.8). Then pointed 
subsequences (M, $g_{i}, p_{i})$ converge to one of the following:

\noindent
$\bullet$ complete cusps $(N, g_{\infty}, p_{\infty})$. These are 
complete, open Riemannian 3-manifolds, of finite volume and with graph 
manifold ends, which collapse at infinity. The convergence is in the 
$C^{1,\alpha}$ and weak $L^{2,p}$ topologies, uniform on compact 
subsets.

\noindent
$\bullet$ Graph manifolds collapsed along local $S^{1}$ or $T^{2}$ actions 
to lower dimensional metric spaces of infinite diameter. The convergence is 
in the Gromov-Hausdorff topology.
\end{theorem}

 In contrast to the topological implications of collapse in Theorem 
3.2, (i.e collapse implies $M$ is a graph manifold), in general there 
are no apriori topological restrictions on $M$ imposed by Theorem 3.3. 
To illustrate, let $M$ be an arbitrary closed 3-manifold and let 
$\{C_{k}\}$ be a collection of disjoint solid tori $D^{2}\times S^{1}$ 
embedded in $M$; for example $\{C_{k}\}$ may be a tubular neighborhood 
of a (possibly trivial) link in $M$. Then it is not difficult to 
construct a sequence of metrics of bounded curvature which converge to 
a collection of complete cusps on $M\setminus \cup C_{k}$ and collapse 
along the standard graph manifold structure on each $C_{k}.$

 The ends of the cusp manifolds $N$ in Theorem 3.3, i.e. the graph 
manifolds, necessarily have embedded tori. If such tori are essential 
in $M$, i.e. inject on the $\pi_{1}$ level, then Theorem 3.3. does 
imply strong topological constraints on the topology of $M$; cf. \S 6 
for some further discussion.

\begin{remark} \label{r 3.4.} 
{\rm We point out that there are versions of Theorems 3.2 and 3.3 also 
in dimension 4, as well as in higher dimensions. The concept of graph 
manifold is generalized to manifolds having an ``F-structure'', or an 
``N-structure'' (F is for flat, N is for nilpotent), cf. [18], provided 
bounds are assumed on the full curvature, as in (1.4). In dimension 4, 
this can be relaxed to bounds on the Ricci curvature, as in (1.5), 
provided one allows for a finite number of singularities in 
F-structure, as in Theorem 2.6. }
\end{remark}

\section{Applications to Static and Stationary Space-Times.}
\setcounter{equation}{0}

 In this section, we discuss applications of the results of \S 2-3 to 
static and stationary space-times, i.e. space-times ({\bf M, g}) which 
admit a time-like Killing field $K$. These space-times are viewed as 
being the end or final state of evolution of a (time dependent) 
gravitational field. Since they are time-independent in a natural 
sense, they may be analysed by methods of Riemannian geometry, which 
are not available in general for Lorentzian manifolds.

 Throughout this section, we assume that ({\bf M, g}) is chronological, 
i.e. ({\bf M, g}) has no closed time-like curves, and that $K$ is a 
complete vector field.

\medskip

 Let $\Sigma $ be the orbit space of the isometric ${\mathbb R}$-action 
generated by the Killing field $K$, and let $\pi: {\bf M} \rightarrow 
\Sigma$ be the projection to the orbit space. The 4-metric {\bf g} has 
the form
\begin{equation} \label{e4.1}
{\bf g} = - u^{2}(dt+\theta )^{2} + \pi^{*}(g), 
\end{equation}
where $K = \partial /\partial t, \theta $ is a connection 1-form for 
the bundle $\pi$, $u^{2} = -{\bf g}(K,K) > 0$ and $g = g_{\Sigma}$ is 
the metric induced on the orbit space.

 The vacuum Einstein equations are equivalent to an elliptic system of 
P.D.E's in the data $(\Sigma, g, u, \theta)$. Let $\omega$ be the twist 
1-form on $\Sigma$, 
given by $2\omega  = *(\kappa\wedge d\kappa) = -u^{4}*d\theta$, where 
$\kappa  = - u^{2}(dt+\theta )$ is the 1-form dual to $K$. (The first $*$ 
operator is on {\bf M} while the second is on $\Sigma$). Then the 
equations on $\Sigma $ are:
\begin{equation} \label{e4.2}
Ric_{g} = u^{-1}D^{2}u + 2u^{-4}(\omega\otimes\omega  -  
|\omega|^{2}g), 
\end{equation}
\begin{equation} \label{e4.3}
\Delta u = - 2u^{-3}|\omega|^{2}, 
\end{equation}
\begin{equation} \label{e4.4}
d\omega  = 0. 
\end{equation}

 The maximum principle applied to (4.3) immediately implies that if 
$\Sigma $ is a closed $3$-manifold, then $(\Sigma, g)$ is flat and $u = 
const$, and so 
({\bf M, g}) is a (space-like) isometric quotient of empty Minkowski 
space 
$({\mathbb R}^{4}, \eta)$. Thus, we assume $\Sigma$ is open, possibly 
with boundary. 

  Locally of course there are many solutions to the system (4.2)-(4.4); 
to obtain uniqueness, one needs to impose boundary conditions.

\bigskip

  We consider first the global situation, and so assume that $(\Sigma, 
g)$ is a complete, non-compact Riemannian 3-manifold. Boundary 
conditions are then at infinity, i.e. conditions on the asymptotic 
behavior of the metric. In this respect, one has the following 
classical result, cf. [37], [24].

\begin{theorem} \label{t 4.1} {\bf (Lichnerowicz).}
The only complete, stationary vacuum space-time ({\bf M, g}) which is 
asymptotically flat (AF) is empty Minkowski space-time $({\mathbb 
R}^{4}, \eta).$
\end{theorem}

 It is most always taken for granted that $\Sigma $ should be AF. 
Stationary space-times are meant to model isolated physical systems, 
and the only physically reasonable models are AF, since in the far-field 
regime, general relativity should approximate Newtonian gravity. In fact, 
from this physical perspective, the Lichnerowicz theorem\index{Lichnerowicz theorem} 
may be viewed as a triviality. Since there is no source for the gravitational field, 
it must be the empty Minkowski space-time. 

 However, mathematically, the Lichnerowicz theorem is not (so) trivial. 
Moreover, the assumption that ({\bf M, g}) is AF is contrary to the 
spirit of general relativity. Such a boundary condition is adhoc, and 
its imposition is mathematically circular in a certain sense. Apriori, there 
might well be complete stationary solutions for which the curvature 
does not decay anywhere to 0 at infinity. From this more general perspective, 
one should be able to {\it deduce} that the far-field regime of stationary 
space-times is necessarily AF and not have to assume this to begin with.

  The following result from [6] clarifies this issue.

\begin{theorem} \label{t 4.2} {\bf (Generalized Lichnerowicz).} 
The only complete stationary vacuum space-time ({\bf M, g}) is empty 
Minkowski space-time $({\mathbb R}^{4}, \eta)$, or a discrete isometric 
quotient of it.
\end{theorem}

 The starting point of the proof of this result is to study first the 
moduli space of all complete stationary vacuum solutions. As noted 
above, any given solution may, apriori, have unbounded 
curvature, i.e. $|Ric_{g}|$ may diverge to infinity on divergent 
sequences in $\Sigma .$ Under such a condition, the first step is then 
to show, by taking suitable base points and rescalings, that one may 
obtain a new stationary vacuum solution, (i.e. a new point in the 
moduli space), with uniformly bounded curvature, and non-zero curvature 
at a base point. This step uses the Cheeger-Gromov theory, as described 
in \S 2-\S 3, and requires the special features of collapse in 
3-dimensions.

 The next step in the proof is to recast the problem in the Ernst 
formulation. Define the conformally related metric $\widetilde g$ by
\begin{equation} \label{e4.5}
\widetilde g = u^{2}g. 
\end{equation}
A simple calculation shows that (4.2) becomes
\begin{equation} \label{e4.6}
Ric_{\widetilde g} =2(d\ln u)^{2} + 2u^{-4}\omega^{2} \geq  0. 
\end{equation}
Further, the system (4.2)-(4.4) becomes the Euler-Lagrange equations 
for an effective 3-dimensional action given by
$${\mathcal S}_{\rm eff} = \int[R - \frac{1}{2}(\frac{|d\phi|^{2} + 
|du^{2}|^{2}}{u^{4}})]dV.$$
Here $\phi$ is the twist potential, given by $d\phi = 2\omega$. (In 
general one must pass to the universal cover to obtain the existence of 
$\phi$). 

  This action is exactly 3-dimensional (Riemannian) gravity on 
$(\Sigma, \widetilde g)$ coupled to a $\sigma$-model with target the 
hyperbolic plane $(H^{2}(-1), g_{-1}).$ Thus, the Ernst map $E = (\phi 
, u^{2})$ is a harmonic map
\begin{equation} \label{e4.7}
E: (\Sigma , \widetilde g) \rightarrow  (H^{2}(-1), g_{-1}). 
\end{equation}

 Now it is well-known that harmonic maps $E: (M, g) \rightarrow  (N, 
h)$ from Riemannian manifolds of non-negative Ricci curvature to 
manifolds of non-positive sectional curvature have strong rigidity 
properties, via the Bochner-Lichnerowicz formula,
\begin{equation} \label{e4.8}
\frac{1}{2}\Delta|DE|^{2} = |D^{2}E|^{2} + \langle Ric_{g}, E^{*}(h) \rangle 
-  \sum (E^{*}R_{h})(e_{i},e_{j},e_{j},e_{i}). 
\end{equation}
By analysing (4.8) carefully, one shows that $E$ is a constant map, 
from which it follows easily that ({\bf M, g}) is flat.

\begin{remark} \label{r 4.3} {\bf (i).}
{\rm The same result and proof holds for stationary gravitational 
fields coupled to $\sigma$-models, whose target spaces are Riemannian 
manifolds of non-positive sectional curvature, i.e. $E: (\Sigma, 
\widetilde g) \rightarrow (N, g_{N})$ with $Riem_{g_{N}} \leq 0$.

  {\bf (ii).} Curiously, the Riemannian analogue of Theorem 4.2 remains 
an open problem. Thus, does there exist a complete non-flat Ricci-flat 
Riemannian 4-manifold which admits a free isometric $S^{1}$ action? 

  {\bf (iii).} It is interesting to note that the analogue of Theorem 
4.2 is false for stationary Einstein-Maxwell solutions. A 
counterexample is provided by the (static) Melvin magnetic universe 
[39], cf. also [28]. I am grateful to David Garfinkle for pointing this 
out to me. For the stationary Einstein-Maxwell system, the target space 
of the Ernst map is $SU(2,1)/S(U(1,1)\times U(1))$, ($SO(2,1)/SO(1,1)$ 
for static Einstein-Maxwell). Both of these target spaces have 
indefinite, (i.e. non-Riemannian), metrics. }
\end{remark}

 The rigidity result Theorem 4.2 leads to apriori estimates on the 
geometry of general stationary solutions of the Einstein equations. 
Thus, if $\Sigma $ is not complete, it follows that 
$\partial\Sigma  \neq  \emptyset$. Note that part of $\partial\Sigma 
$ may correspond to the horizon $H = \{u =$ 0\} where the Killing field 
vanishes. The following result is also from [6].

\begin{theorem} \label{t 4.4} {\bf (Curvature Estimate).} \index{curvature 
estimate} 
Let ({\bf M, g}) be a stationary vacuum space-time. Then there is a 
constant 
$C <  \infty ,$ independent of ({\bf M, g}), such that
\begin{equation} \label{e4.9}
|{\bf R}|(x) \leq  C/r^{2}[x], 
\end{equation}
where $r[x] = dist_{\Sigma}(\pi (x), \partial\Sigma ).$ 
\end{theorem}

 Here, the curvature norm $|{\bf R}|$ may be given by 
$$|{\bf R}|^{2} = |R_{\Sigma}|^{2} + |d\ln u|^{2} + 
|u^{-2}\omega|^{2}.$$ 
Note that Theorem 4.2 follows from Theorem 4.4 by letting $r 
\rightarrow \infty$. Conversely, it is a general principle for elliptic 
geometric variational problems that a global rigidity result as in 
Theorem 4.2 leads to apriori local estimates as in Theorem 4.4.

\begin{remark} \label{r 4.5} {\bf (i).}
{\rm Using elliptic regularity, one also has higher order bounds:
\begin{equation} \label{e4.10}
|\nabla^{k}{\bf R}|(x) \leq  C_{k}/r^{2+k}[x]. 
\end{equation}

 {\bf (ii).} A version of this result also holds for stationary 
space-times with energy-momentum tensor $T$. Thus, for example one has
\begin{equation} \label{e4.11}
|{\bf R}
|(x) \leq  C_{\alpha}\cdot |T|_{C^{\alpha}(B_{[x]}(1))}, 
\end{equation}
for any $\alpha > 0$, where $B_{[x]}(1)$ is the unit ball in $(\Sigma, 
g)$ about $[x]$. The proof is the same as that of (4.9) given in [6]. }
\end{remark}

 Thus, one can use the Cheeger-Gromov theory to control the local 
behavior of stationary space-times, possibly with matter terms, away 
from any boundary.

\medskip

 The results above can in turn be applied to study the possible 
asymptotic behavior of general stationary or static vacuum space-times, 
without any apriori AF assumption. For example, (4.9) implies that the 
curvature decays at least quadratically in any end $(E, g)$ of 
$(\Sigma, g)$. For simplicity, we restrict here to static space-times. 

  Thus, let ({\bf M, g}) be a static space-time with orbit space 
$(\Sigma, g)$, with $\partial\Sigma  \neq  \emptyset$. Define 
$\partial\Sigma$ to be {\it  pseudo-compact} if there exists $r_{o} > 
0$ such that the level set $\{r  = r_{o}\}$ in $\Sigma $ is compact; 
recall that $r$ is the distance function to the boundary $\partial 
\Sigma$. (There are numerous examples of static space-times for which 
$\partial\Sigma $ is non-compact, with $\partial \Sigma$ 
pseudo-compact). Let $S(s) = r^{-1}(s) \subset  \Sigma$. If $E$ is an 
end of $(\Sigma, g)$, define its mass $m_{E}$ by
\begin{equation} \label{e4.12}
m_{E} = \lim_{s\rightarrow\infty}\frac{1}{4\pi}\int_{S(s)} 
\langle \nabla \ln u, \nabla r \rangle dA. 
\end{equation}
It is easily seen from the static vacuum equations that the integral is 
monotone non-increasing in $s$, and so the limit exists. The mass 
$m_{E}$ coincides with the Komar mass in case $E$ is AF. The following 
result is from [7].

\begin{theorem} \label{t 4.6} {\bf (Static Asymptotics).} 
Let ({\bf M, g}, u) be a static vacuum space-time with pseudo-compact 
boundary. Then 
({\bf M, g}) has a finite number of ends. Any end $E$ on which
\begin{equation} \label{e4.13}
\liminf_{E}u >  0, 
\end{equation}
is either:
\begin{center}
AF
\end{center}
or
\begin{equation} \label{e4.14}
{\rm small} \equiv_{def} \int_{1}^{\infty}[areaS(r)]^{-1}dr <  \infty . 
\end{equation}
Further, if $m_{E} \neq $ 0 and $sup_{E}u <  \infty ,$ then $E$ is AF. 
\end{theorem}

 This result is sharp in the sense that if any of the hypotheses are 
dropped, then the conclusion is false. For instance, if (4.13) fails, 
then there are examples of static vacuum solutions with ends neither 
small nor AF.

 We note that when $E$ is AF, the result implies it is AF in the strong 
sense that
\begin{equation} \label{e4.15}
|g -  g_{0}| = \frac{2m}{r} + O(r^{-2}), \ |R| = O(r^{-3}), \ {\rm and} 
\ |u- 1| = \frac{m}{r} + O(r^{-2}). 
\end{equation}
More precise asymptotics can then be obtained by using standard 
elliptic estimates on the equations (4.2)-(4.4), or from [14]. Again, a 
version of Theorem 4.6 holds for static space-times with matter, cf. 
again [7] for further information.

 The idea of the proof is to study the asymptotic behavior of an end 
$E$ by blowing it down, as described in \S 1. Thus, for $R$ large and 
any fixed $k$, consider the metric annuli $A(R, kR)$ about some base 
point $x_{o}\in (\Sigma, g)$ and consider the rescalings $g_{R} = 
R^{-2}g.$ The annulus $A(R, kR)$ then becomes an annulus of the metric 
form $A(1,k)$ w.r.t. $g_{R}.$ Further, the estimate (4.9) implies that 
the curvature of $g_{R}$ in $A(1,k)$ is uniformly bounded. Thus, one 
may apply the Cheeger-Gromov theory as described in \S 2,\S 3, to a 
sequence $(A(1,k), g_{R_{i}})$, with $R_{i} \rightarrow  \infty$. One 
proves that the convergence case gives rise to AF ends, while the 
collapse case gives rise to small ends.

\medskip

 Note that in the collapsing situation, one obtains an extra $S^{1}$ or 
$T^{2}$ symmetry when the collapse is unwrapped in covering spaces. 
Thus, the behavior in this case is described by axisymmetric static 
solutions, i.e. the Weyl metrics. Small ends typically have the same 
end structure as ${\mathbb R}^{2}\times S^{1},$ where the $S^{1}$ 
factor has bounded length and so typically have at most quadratic 
growth for the area of geodesic spheres. 

\medskip

 It is worth pointing out that there are static vacuum solutions, 
smooth up to a compact horizon, which have a single small end. This is 
the family of Myers metrics [40], or periodic Schwarzschild metrics, 
(discovered later and independently by Korotkin and Nicolai). The 
manifold $\Sigma $ is topologically $(D^{2}\times S^{1})\setminus 
B^{3},$ so that $\partial\Sigma  = S^{2}$ with a single end of the form 
$T^{2}\times {\mathbb R}^{+}.$ Metrically, the end is asymptotic to one 
of the (static) Kasner metrics. This is of course not a counterexample 
to the static black hole uniqueness theorem, since the end is not AF.

  Note that since $\pi_{1}(\Sigma) = {\mathbb Z}$ here, one may take 
non-trivial covering spaces of the Myers metrics. This leads to static 
vacuum solutions with an arbitrary finite number, or even an infinite 
number, of black holes in static equilibrium. This situation is of 
course not possible in Newtonian gravity, and so is a highly non-linear 
effect of general relativity.

\section{Lorentzian Analogues and Open Problems.}
\setcounter{equation}{0}

 In this section, we discuss potential analogues of the results of \S 2 
and \S 3 for Lorentzian metrics on 4-manifolds. The main interest is in 
space-times ({\bf M, g}) for which one has control on the Ricci 
curvature of {\bf g}, or via the Einstein equations, control on the 
energy-momentum tensor $T$. In particular, the main focus will be on 
vacuum space-times, $Ric_{{\bf g}} = 0$.

\medskip

 One would like to find conditions under which one can take limits of 
vacuum space-times. One natural reason for trying to do this is the 
following. There are now a number of situations where global stability 
results have been proved, namely: the global stability of Minkowski 
space-time [21], and of deSitter space-time [26], the global future 
stability of the Milne space-time [10], and the future $U(1)$ stability 
of certain Bianchi models [20]. These results are {\it  openness}  
results, which state that the basic features of a given model, e.g. 
Minkowski, are preserved under suitably small perturbations of the 
initial data. It is then natural to consider what occurs when one tries 
to pass to limits of such perturbations.

 The issue of being able to take limits is also closely related with 
the existence problem and singularity formation for the vacuum Einstein 
evolution equations. From this perspective, suppose one has an 
increasing sequence of domains $(\Omega_{i}, {\bf g_{i}})$, $\Omega_{i} 
\subset  \Omega_{i+1}$ with ${\bf g_{i+1}}|_{\Omega_{i}} = {\bf g_{i}}$, 
which are evolutions of smooth Cauchy data on some fixed initial data set. 
If ${\bf M} = \cup\Omega_{i}$ is the maximal Cauchy development, then 
understanding {\bf (M, g)} amounts to understanding the limiting behavior of 
$(\Omega_{i}, {\bf g_{i}}).$

\medskip

 There are two obvious but essential reasons why it is much more 
difficult to develop a Lorentzian analogue of the Cheeger-Gromov 
theory, in particular with bounds only on the Ricci curvature. The 
first is that the elliptic nature of the P.D.E. for Ricci curvature 
becomes hyperbolic for Lorentz metrics, and hyperbolic P.D.E. are much 
more difficult than elliptic P.D.E. The second is that the group of 
Euclidean rotations $O(4)$ is compact, while the group of proper Lorentz 
transformations $O(3,1)$ is non-compact.

\medskip
\noindent
{\bf A: $1^{\rm st}$ Level Problem.}
 
 Consider first the problem of controlling the space-time metric {\bf 
g} in terms of bounds, say $L^{\infty},$ on the space-time curvature 
{\bf R}, 
\begin{equation} \label{e5.1}
|{\bf R}|_{L^{\infty}} \leq K < \infty,
\end{equation}
since already here there are significant issues.

 First, the norm of curvature tensor 
$|{\bf R}|^{2} = {\bf R}_{ijkl}{\bf R}^{ijkl}$ is no longer 
non-negative for 
Lorentz metrics, and so a bound on $|{\bf R}|^{2}$ does not imply a 
bound on 
all the components ${\bf R}_{ijkl}.$ In fact, for a Ricci-flat 
4-metric, 
there are exactly two scalar invariants of the curvature tensor:
\begin{equation} \label{e5.2}
\langle {\bf R}, {\bf R} \rangle = |{\bf R}|^{2} = {\bf R}_{ijkl}{\bf R}^{ijkl} \ 
{\rm and} \ 
\langle {\bf R}, *{\bf R} \rangle = {\bf R}_{ijkl}(*{\bf R}^{ijkl}). 
\end{equation}
Both of these invariants can vanish identically on classes of 
Ricci-flat non-flat space-times; for instance this is the case for the 
class of plane-fronted gravitational waves, given by
$${\bf g} = -dudv + (dx^{2} + dy^{2}) - 2h(u,x,y)du^{2},$$
$$\Delta_{(x,y)}h = 0,$$
cf. [15,\S 8] and references therein. Here, $h$ is only required to 
harmonic in the variables $(x,y)$, and is arbitrary in $u$. The class 
of such space-times is highly non-compact, and so one has no local 
control of the metric in any coordinate system under bounds on the 
quantities in (5.2).

\medskip

 Thus, one must turn to bounds on the components of {\bf R} in some 
fixed coordinate system or framing. The most efficient way to do this 
is to choose a unit time-like vector $T = e_{0},$ say future directed, 
and extend it to an orthornormal frame $e_{\alpha},$ 0 $\leq  \alpha  
\leq $ 3. Since the space $T^{\perp}$ orthogonal to $T$ is space-like 
and $O(3)$ is compact, the particular framing of $T^{\perp}$ is 
unimportant. One may then define the norm w.r.t. $T$ by
\begin{equation} \label{e5.3}
|{\bf R}|_{T}^{2} = \sum ({\bf R}_{ijkl})^{2}, 
\end{equation}
where the components are w.r.t. the frame $e_{\alpha}.$ This is 
equivalent to taking the norm of {\bf R} w.r.t. the Riemannian metric 
$$g_{E} = {\bf g} + 2T\otimes T.$$ 

 If, at a given point $p$, $T$ lies within a compact subset $W$ of the 
future interior null cone $T_{p}^{+}$, then the norms (5.3) are all 
equivalent, with constant depending only on $W$. Of course if $D$ is a 
compact set in the space-time ({\bf M, g}) and the vector field $T$ is 
continuous in $D$, then $T$ lies within a compact subset of $T^{+}D,$ 
where $T^{+}D$ is the bundle of future interior null cones in the 
tangent bundle $TD$.

\medskip

 It is quite straightforward to prove that if $(M, g)$ is a smooth 
Riemannian manifold with an $L^{\infty}$ bound on the full curvature, 
$|R| \leq  K$ then there are local coordinate systems in which the 
metric is $C^{1,\alpha}$ or $L^{2,p},$ with bounds depending only on 
$K$ and a lower volume bound, cf. Remark 2.4(i). 

  However, this has been an open problem for Lorentzian metrics, 
apparently for some time, cf. [22],[47] for instance. The following 
result gives a solution to this problem.

\medskip

  To state the result, we need the following definition. Let $\Omega$ be a 
domain in a smoooth Lorentz manifold ({\bf M, g}), of arbitrary dimension 
$n+1$. Then $\Omega$ is said to satisfy the {\sf size conditions} if the 
following holds. There is a smooth time function $t$, with 
$T = \nabla t / |\nabla t|$ the associated unit time-like vector field on 
$\Omega$, such that, for $S = S_{0} = t^{-1}(0)$, the 1-cylinder
\begin{equation} \label{e5.4}
C_{1} = B_{p}(1) \times [-1, 1] \subset \subset \Omega,
\end{equation}
i.e. $C_{1}$ has compact closure in $\Omega$. Here $B_{p}(1)$ is the 
geodesic ball of radius 1 about $p$, w.r.t. the metric $g$ induced on 
$S$ and the product is identified with a subset of $\Omega$ by the flow 
of $T$. 

  It is essentially obvious that any point $q$ in a Lorentz manifold 
has a neighborhood satisfying the size conditions, when the metric {\bf 
g} is scaled up suitably.

  Let $D = Im T|_{C_{1}} \subset \subset T^{+}\Omega$.
\begin{theorem} \label{t 5.1}
Let $\Omega$ be a domain in a vacuum $(n+1)$-dimensional space-time 
$({\bf M, g})$. Suppose $\Omega$ satisfies the size conditions, and 
that there exist constants $K <  \infty $ and $v_{o} > $ 0 such that
\begin{equation} \label{e5.5}
|{\bf R}|_{T} \leq  K,  \ \ vol_{g}B_{p}(\textstyle{\frac{1}{2}}) \geq  
v_{o}. 
\end{equation}
Then there exists $r_{o} > 0$, depending only on $K, v_{o}$ and $D$, 
and coordinate charts on the $r_{o}$-cylinder
$$C_{r_{o}} = B_{p}(r_{o}) \times [-r_{o}, r_{o}],$$
such that the components of the metric ${\bf g}_{\alpha\beta}$ are in 
$C^{1,\alpha}\cap L^{2,p}$, for any $\alpha < 1$, $p <  \infty$. 

  Further, there exists $R_{o}$, depending only on $K$, $v_{o}$, $D$ 
and $p$, such that, on $C_{r_{o}}$,
\begin{equation} \label{e5.6}
||{\bf g}_{\alpha\beta}||_{L^{2,p}} \leq  R_{o}. 
\end{equation}
\end{theorem}

  Here, the components ${\bf g}_{\alpha\beta}$ are the full space-time 
components of {\bf g}, and the estimate (5.6) gives bounds on both 
spatial and time derivatives of {\bf g}, up to order 2, in $L^{p}$, 
where $L^{p}$ is measured on spatial slices of $C_{r_{o}}$.

 This result is formulated in such a way that it is easy to pass to 
limits. Thus, if one has a sequence of smooth space-times $({\bf 
M}_{i}, {\bf g}_{i})$ satisfying the hypotheses of the Theorem, (with 
fixed constants $K$, $v_{o}$ and uniformly compact domains $D$), then 
it follows that, in a subsequence, there is a limit $C^{1,\alpha}\cap 
L^{2,p}$ space-time $({\bf M_{\infty}, g_{\infty}})$, defined at least 
on the $r_{o}$-cylinder $C_{r_{o}}$. Further, the convergence to the 
limit is $C^{1,\alpha}$ and weak $L^{2,p}$, and the estimate (5.6) 
holds on the limit.

\medskip

 We sketch some of the ideas of the proof; full details appear in [9]. 
First, one constructs a new local time function $\tau$ on small 
cylinders $C_{r_{1}}$, with $|\nabla \tau|^{2} = -1$, so the flow of 
$\nabla \tau$ is by time-like geodesics. On the level sets 
$\Sigma_{\tau}$ of $\tau$, one constructs spatially harmonic 
coordinates $\{x_{i}\}$, (w.r.t. the induced Riemannian metric). This 
gives a local coordinate system $(\tau, x_{1}, ... , x_{n})$ on small 
cylinders about $p$. One then uses the transport or Raychaudhuri 
equation, together with the Bochner-Weitzenbock formula, (Simons' 
equation), and elliptic estimates to control ${\bf g}_{\alpha \beta}$.

  The vacuum Einstein equations are needed in Theorem 5.1 only to prove 
the $2^{\rm nd}$ time derivatives of the shift 
$\partial_{\tau}\partial_{\tau} {\bf g}_{0\alpha}$ are in $L^{p}$, via use of 
the Bianchi identity. In place of vacuum space-times, it suffices to have a 
rather weak bound on the stress-energy tensor in the Einstein equations. 
All other bounds on ${\bf g}_{\alpha\beta}$ do not require the Einstein 
equations.

\medskip

 It would be interesting to apply this result, or variants of it, to obtain 
further information on the structure of the boundary of space-times.

\medskip

  If the volume bound on space-like hypersurfaces in (5.5) is dropped, 
then it is possible that space-like hypersurfaces may collapse with 
bounded curvature, as described in \S 3. Examples of this behavior 
occur on approach to Cauchy horizons, (as noted in \S 3 in connection 
with the Berger collapse and the Taub-NUT metric). More generally, 
Rendall [45] has proved the following interesting general result: if 
$\Sigma$ is a {\it compact} Cauchy horizon in a smooth vacuum 
space-time in 3+1 dimensions, then nearby space-like hypersurfaces 
collapse with bounded curvature on approach to $\Sigma$.

\bigskip
\noindent
{\bf B: $2^{\rm nd}$ Level Problem.} 

  While Theorem 5.1 represents a first step, one would like to do much 
better by replacing the bound on $|{\bf R}|_{T}$ by a bound on the 
Ricci curvature of ({\bf M, g}), or assuming for instance the vacuum 
Einstein equations. Thus, one may ask if analogues of Theorems 2.2 or 
2.3 hold in the Lorentzian setting.

 The main ingredients in the proofs of these results are the splitting 
theorem - a geometric part - and the strong convergence to limits - an 
analytic part obtained from elliptic estimates for the Ricci curvature. 
\index{splitting theorem}
Now one does have a direct analogue of the splitting theorem for vacuum 
space-times, (or more generally space-times satisfying the time-like 
convergence condition). Thus, by work of Eschenburg, Galloway and 
Newman, if ({\bf M, g}) is a time-like geodesically complete, (or a 
globally hyperbolic), vacuum space-time which contains a time-like 
line, i.e. a complete time-like maximal geodesic, then ({\bf M, g}) is 
flat, cf. [13] and references therein.

 In analogy to the Riemannian case, define then the 1-cross 
$Cro_{1}(x,T)$ of a Lorentzian 4-manifold ({\bf M, g}) at $x$, in the 
direction of a unit time-like vector $T$, to be the length of the 
longest maximizing geodesic in the direction $T$, with center point 
$x$. For $\Omega$ a domain with compact closure in 
{\bf M} and $T$ a smooth unit time-like vector field, define 
$$Cro_{1}(\Omega ,T) = \inf_{x\in\Omega}Cro_{1}(x,T).$$

 What is lacking is the regularity boost obtained from elliptic 
estimates. For space-times, the vacuum equations give a hyperbolic 
evolution equation, (in harmonic coordinates), for which one does not 
have a gain in derivatives. However, the smoothness of initial data is 
preserved under the evolution, until one hits the boundary of the 
maximal development.

 Let $H^{s} = H^{s}(U)$ denote the Sobolev space of functions with $s$ 
weak derivatives in $L^{2}(U)$, $U$ a bounded domain in ${\mathbb 
R}^{3}$. For $s > 2.5$, (so that $H^{s}$ embeds in $C^{1}$), and a 
space-like hypersurface $S \subset $ ({\bf M, g}), define the harmonic 
radius \index{harmonic radius} $\rho_{s}(x)$ of $x\in S$ in the same way 
as in Definition 2.1, where the components ${\bf g}_{\alpha \beta}$ and 
derivatives are in both space and time directions. For the following, we 
need only consider $s \in {\mathbb N}^{+}$, with $s$ large, for 
instance, $s = 3$. 

  Now a well-known result of Choquet-Bruhat [19] states that the 
maximal vacuum $H^{s}$ development of smooth ($C^{\infty}$) initial 
data on $S$ is the same for all $s$, provided $s > 2.5$. Thus, one does 
not have different developments of smooth initial data, depending on 
the degree of desired $H^{s}$ regularity. Here, one may assume that $S$ 
is compact, or work locally, within the domain of dependence of $S$. 
This qualitative result can be expressed as follows. Let $S_{t}$ be 
space-like hypersurfaces obtained by evolution from initial data on $S 
= S_{0}$. If $x_{t} \in S_{t}$, then
\begin{equation} \label{e5.7}
\rho_{s}(x_{t}) \geq  c_{1} \Rightarrow  \rho_{s+1}(x_{t}) \geq  c_{2}, 
\end{equation}
where $c_{2}$ depends on $c_{1}$ and the $(C^{\infty})$ initial data on 
$S_{0}$.

 We raise the following problem of whether the qualitative statement 
(5.7) can be improved to a {\it quantitative} statement.

{\bf Regularity Problem.}
 Can the estimate (5.7) be improved to an estimate
\begin{equation} \label{e5.8}
\inf_{x_{t} \in S_{t}}\rho_{s+1}(x_{t}) \geq  c_{0}\cdot \inf_{x_{t} \in 
S_{t}}\rho_{s}(x_{t}), 
\end{equation}
where $c_{0}$ depends only on the initial data on $S$? One may assume, 
w.l.o.g, that $t \leq 1$.

 The important point of (5.8) over (5.7) is that the estimate (5.8) is 
scale-invariant. Here, we recall that $\rho_{s}(x)$ measures the degree 
of concentration of derivatives of the metric in $H^{s}$, so that 
$\rho_{s} \rightarrow 0$ corresponds to blow-up of the metric in 
$H^{s}$ locally. 

 If (5.8) holds, it serves as an analogue of the regularity boost. In 
such circumstances, one can imitate the proof of Theorems 2.2 or 2.3 to 
obtain similar results for sequences of space-times $({\bf M, g}_{i})$. 

 In fact, the validity of (5.8) would have numerous interesting 
applications, even if it could be established under some further 
restrictions or assumptions.

\bigskip

  Suppose next one drops any assumption on the 1-cross of ({\bf M, g}) 
and maintains only a lower bound on the volumes of geodesic balls, as 
in (5.5), on space-like hypersurfaces. This leads directly to issues of 
singularity formation and the structure of the boundary of the vacuum 
space-time, where comparatively little is known mathematically.

  A useful problem, certainly simple to state, is the following: for 
simplicity, we work in the context of compact, (i.e. closed, without 
boundary), Cauchy surfaces.

\smallskip
\noindent
{\bf Sandwich Problem.}

 Let $({\bf M, g}_{i})$ be a sequence of vacuum space-times, and let 
$\Sigma_{i}^{1}, \Sigma_{i}^{2}$ be two compact Cauchy surfaces in 
${\bf M}$, with $\Sigma_{i}^{2}$ to the future of $\Sigma_{i}^{1}$ and 
with 
$$1 \leq dist_{{\bf g}}(x, \Sigma_{i}^{1}) \leq 10,$$
for all $x \in \Sigma_{i}^{2}$. Suppose the Cauchy data $(g_{i}^{j}, 
K_{i}^{j})$, $j = 1,2$ on each Cauchy surface are uniformly bounded in 
$H^{s}$ for some fixed $s > 2.5$, possibly large. Hence the data 
$(g_{i}^{j}, K_{i}^{j})$ converge, in a subsequence and weakly in 
$H^{s}$, to limit $H^{s}$ Cauchy data $g_{\infty}^{j}, K_{\infty}^{j}$ 
on $\Sigma^{j}$. 

  Do the vacuum space-times $A_{i}(1,2) \subset (M, g_{i})$ between 
$\Sigma^{1}$ and $\Sigma^{2}$ converge, weakly in $H^{s}$, to a limit 
space time,
\begin{equation} \label {e5.9}
(A_{i}(1,2), g_{i}) \rightarrow  (A_{\infty}, g_{\infty})?
\end{equation}
This question basically asks if a singularity can form between 
$\Sigma_{i}^{1}$ and $\Sigma_{i}^{2}$ in the limit. It is unknown even 
if there could be only a single singularity at an isolated point 
(event) $x_{0} \in (A_{\infty}, g_{\infty})$. 

  The existence of such a singularity may be related to the Choptuik 
solution. \index{Choptuik solution} However, both the existence and the 
smoothness properties of the Choptuik solution have not been established 
well mathematically; cf. [34] for an interesting discussion.

  Such a limit singularity would be naked in a strange way. It could be 
detected on $\Sigma^{2}$, since light rays from it propagate to 
$\Sigma^{2}$, but on $\Sigma^{2}$, no remnant of the singularity is 
detectable, since the data is smooth on $\Sigma^{2}$. Thus, the 
singularity is invisible to the future (or past) in a natural sense.

  A resolution of this problem would be useful in understanding, for 
instance, limits of the asymptotically simple vacuum perturbations of 
deSitter space, given by Friedrich's theorem [26]. The sandwich problem 
above asks: suppose one has control on the space-time near past and 
future space-like infinity $\mathcal{I}^{\pm}$, does it follow that one 
has control in between?

  Similar questions can be posed for non-compact Cauchy surfaces, and 
relate for instance to limits of the AF perturbations of Minkowski 
space given by Christodoulou-Klainerman, [21].

\section{Future Asymptotics and Geometrization of 3-Manifolds.}
\setcounter{equation}{0}

 In this section, we give some applications to the future asymptotic 
behavior of cosmological spaces times.

 Let ({\bf M, g}) be a vacuum cosmological space-time, i.e. ({\bf M, g}) 
contains a compact Cauchy surface $\Sigma$ of constant mean 
curvature (CMC). It is well-known that $\Sigma$ then embeds 
in a (local) foliation $\mathcal{F}$ by CMC Cauchy surfaces $\Sigma_{\tau}$, 
all diffeomorphic to $\Sigma  = \Sigma_{1}$, and parametrized by their 
mean curvature $\tau$. The parameter $\tau$ thus serves as a time 
function, with respect to which one may describe the evolution of the 
space-time. We refer to the work of Bartnik [11], [12] and Gerhardt [29] for 
results on the existence of such foliations, and to the surveys by 
Marsden-Tipler [38] and Rendall [44] for an overview of this topic.

  We assume throughout this section that $\Sigma$ is of non-positive 
Yamabe type, i.e. $\Sigma$ admits no metric of positive scalar curvature. 
It then follows from the Hamiltonian constraint equation that the mean 
curvature $\tau$ never achieves the value $0$. Thus 
\begin{equation} \label{e6.1}
\tau  \in  (-\infty , 0), 
\end{equation}
with $\tau$ increasing towards the future in ({\bf M, g}). The sign of 
the mean curvature is chosen so that $vol_{g_{\tau}}\Sigma_{\tau}$ is 
increasing with increasing $\tau$, i.e. expanding towards the future. 
The foliated region ${\bf M}_{\mathcal{F}}$ is thus a subset of {\bf M}, 
although in general one cannot expect that ${\bf M} = {\bf M}_{\mathcal{F}}$ 
due to the formation of singularities. 

\medskip

 Suppose that ({\bf M, g}) is geodesically complete to the future of 
$\Sigma ,$ and that the future is foliated by CMC Cauchy surfaces, i.e. 
${\bf M} = {\bf M}_{\mathcal{F}}$ to the future of $\Sigma$. These are 
of course strong assumptions, but are necessary if one wants to 
understand the future asymptotic behavior of ({\bf M, g}) without the 
complicating issue of singularities.

 The topology of $\Sigma_{\tau}$ is fixed, and so the metrics 
$g_{\tau}$ induced on $\Sigma_{\tau}$ by the ambient metric {\bf g} 
give rise to a curve of Riemannian metrics on the fixed manifold 
$\Sigma$. In all known situations, one has $vol_{g_{\tau}}\Sigma  
\rightarrow  \infty$ as $\tau  \rightarrow 0$, and the metrics 
$g_{\tau}$ become locally flat, due to the expansion, compare with the 
discussion in \S 1. It would be of interest to prove these statements in 
general, although it is hard to imagine situations where either one of them 
fails.

 The local geometry of $g_{\tau}$ thus becomes trivial locally. This is of 
course not very interesting. As in \S 1 and \S 4, to study 
the asymptotic behavior, one should rescale by the distance to a fixed 
base point or space-like hypersurface. In this case, the distance is 
the time-like Lorentzian distance. Thus, for $x$ to the future of 
$\Sigma  = \Sigma_{-1},$ let $t(x) = dist_{{\bf g}}(x, \Sigma )$ and let
\begin{equation} \label{e6.2}
t_{\tau} = t_{max}(\tau ) = max\{t(x): x\in\Sigma_{\tau}\} = dist_{{\bf 
g}}(\Sigma_{\tau}, \Sigma). 
\end{equation}
It is natural to study the asymptotic behavior of the metrics
\begin{equation} \label{e6.3}
\bar g_{\tau} = t_{\tau}^{-2}g_{\tau}, 
\end{equation}
on $\Sigma_{\tau}.$ Observe that in the rescaled space-time (M, ${\bf 
\bar g}_{\tau}),$ the distance of $(\Sigma_{\tau}, \bar g_{\tau})$ to 
the ``initial'' singularity, (big bang), tends towards 1, as $\tau  
\rightarrow $ 0. Any other essentially distinct scaling would have the 
property that the distance to the initial singularity tends towards 0 
or $\infty$, and so is not particularly natural.

\smallskip

 We need the following definition, closely related to Thurston's 
Geometrization Conjecture \index{geometrization conjecture} [46] on the 
structure of 3-manifolds.

\begin{definition} \label{d 6.1}
{\rm Let $\Sigma $ be a closed, oriented, connected 3-manifold, of 
non-positive Yamabe type. A {\it weak}  geometrization of $\Sigma $ is 
a decomposition
\begin{equation} \label{e6.4}
\Sigma  = H \cup  G, 
\end{equation}
where $H$ is a finite collection of complete, connected hyperbolic 
manifolds, of finite volume, embedded in $\Sigma ,$ and $G$ is a finite 
collection of connected graph manifolds, embedded in $\Sigma .$ The 
union is along a finite collection of embedded tori $\mathcal{T}  = 
\cup T_{i} = \partial H = \partial G.$

 A {\it  strong}  geometrization of $\Sigma $ is a weak geometrization 
as above, for which each torus $T_{i}\in \mathcal{T} $ is 
incompressible in $\Sigma ,$ i.e. the inclusion of $T_{i}$ into $\Sigma 
$ induces an injection of fundamental groups.}
\end{definition}

 Of course it is possible that the collection $\mathcal{T} $ of tori 
dividing $H$ and $G$ is empty, in which case weak and strong 
geometrizations coincide. In such a situation, $\Sigma $ is then either 
a closed hyperbolic manifold \index{hyperbolic manifold} or a closed graph 
manifold. For a strong geometrization, the decomposition (6.4) is unique up 
to isotopy, but this is certainly not the case for a weak geometrization, 
c.f. the end of \S 3.

\medskip

 In general, no fixed metric $g$ on $\Sigma $ will realize the 
decomposition (6.4), unless $\mathcal{T}  = \emptyset$. This is because 
the complete hyperbolic metric on $H$ does not extend to a metric on 
$\Sigma$. However, one can find sequences of metrics $g_{i}$ on $\Sigma 
$ which limit on a geometrization of $\Sigma $ in the sense of (6.4). 
Thus, metrics $g_{i}$ may be chosen to converge to the hyperbolic 
metric on larger and larger compact subsets of $H$, to be more and more 
collapsed with bounded curvature on $G$, and such that their behavior 
matches far down the collapsing hyperbolic cusps.

\medskip

 Next, to proceed further, we need to impose a rather strong curvature 
assumption on the ambient space-time curvature. Thus, suppose there is 
a constant $C <  \infty $ such that, for $x$ to the future of $\Sigma ,$
\begin{equation} \label{e6.5}
|{\bf R}|(x) + t(x)|\nabla{\bf R}|(x) \leq  C\cdot  t^{-2}(x). 
\end{equation}
Here, the curvature norm $|{\bf R}|$ may be given by $|{\bf R}|_{T}$ as 
in (5.3), where $T$ is the unit normal to the foliation 
$\Sigma_{\tau}$. Since ({\bf M, g}) is vacuum, this is equivalent to 
$|{\bf R}|^{2} = |E|^{2} + |B|^{2}$, where $E$, $B$ is the 
electric/magnetic decomposition of {\bf R}, 
$E(X,Y) = \langle {\bf R}(X,T)T, Y \rangle$, 
$B(X,Y) = \langle (*{\bf R})(X,T)T, Y \rangle$ with 
$X$, $Y$ tangent to the leaves. Similarly, $|\nabla{\bf R}|^{2} = 
|\nabla E|^{2} + |\nabla B|^{2}.$ 

 The bound (6.5) is scale invariant, and analogous to the bound (4.9) 
or (4.10) for stationary space-times, (where it of course holds in 
general). The bound on $|\nabla {\bf R}|$ in (6.5) is needed only for 
technical reasons, (related to Cauchy stability), and may be removed in 
certain natural situations.

\medskip

 The discussion above leads to the following result from [8], to which 
we refer for further discussion and details.

\begin{theorem} \label{t 6.2}
Let ({\bf M, g}) be a cosmological space-time of non-positive Yamabe 
type. Suppose that the curvature assumption (6.5) holds, and that 
${\bf M}_{\mathcal{F}} = {\bf M}$ to the future of $\Sigma$.

 Then ({\bf M, g}) is future geodesically complete and, for any 
sequence $\tau_{i} \rightarrow $ 0, the slices $(\Sigma_{\tau_{i}}, 
\bar g_{\tau_{i}})$, cf. (6.3), have a subsequence converging to a weak 
geometrization of $\Sigma ,$ in the sense following Definition 6.1.
\end{theorem}

 We indicate some of the basic ideas in the proof. The first step is to 
show that the bound (6.5) on the ambient curvature {\bf R}, in this 
rescaling, gives uniform bounds on the intrinsic and extrinsic 
curvature of the leaves $\Sigma_{\tau}.$ The proof of this is similar 
to the proof of Theorem 5.1.

 Given this, one can then apply the Cheeger-Gromov theory, as described 
in \S 2- \S 3. Given any sequence $\tau_{i} \rightarrow 0$, there exist 
subsequences which either converge, collapse or form cusps. From the 
work in \S 3, one knows that the regions of $(\Sigma_{\tau_{i}}, \bar 
g_{\tau_{i}})$ which (fully) collapse, or which are sufficiently 
collapsed, are graph manifolds. This gives rise to the region $G$ in 
(6.4). It remains to show that, for any fixed $\varepsilon > 0$, the 
$\varepsilon$-thick region $\Sigma^{\varepsilon}$ of 
$(\Sigma_{\tau_{i}}, \bar g_{\tau_{i}})$ converges to a hyperbolic 
metric.

 The main ingredient in this is the following volume monotonicity 
result:
\begin{equation} \label{e6.6}
\frac{vol_{g_{\tau}} \Sigma_{\tau}}{t_{\tau}^{3}} \downarrow , 
\end{equation}
i.e. the ratio is monotone non-increasing in the distance $t_{\tau}.$ 
This result is analogous to the Fischer-Moncrief monotonicity of the 
reduced Hamiltonian along the CMC Einstein flow, cf. [25]. The 
monotonicity (6.6) is easy to prove, and is an analogue of the 
Bishop-Gromov volume monotonicity (2.12). It follows from an analysis 
of the Raychaudhuri equation, much as in the Penrose-Hawking 
singularity theorems, together with a standard maximum principle.

 Moreover, the ratio in (6.6) is constant on some interval $[\tau_{1}, 
\tau_{2}]$ if and only if the annular region $\tau^{-1}(\tau_{1}, 
\tau_{2})$ is a time annulus in a flat Lorentzian cone
\begin{equation} \label{e6.7}
{\bf g_{o}} = - dt^{2} + t^{2}g_{-1},
\end{equation}
where $g_{-1}$ is a hyperbolic metric. Again, the ratio in (6.6) is 
scale invariant, and so
\begin{equation} \label{e6.8}
\frac{vol_{g_{\tau}} \Sigma_{\tau}}{t_{\tau}^{3}} = vol_{\bar 
g_{\tau}}\Sigma_{\tau}.
\end{equation}
In the non-collapse situation, $vol_{\bar g_{\tau}}\Sigma_{\tau}$ is 
uniformly bounded away from 0 as $\tau \rightarrow 0$, (i.e. $t_{\tau} 
\rightarrow \infty$), and hence converges to a non-zero limit. On 
approach to the $\tau = 0$ limit, the ratio (6.6) tends to a constant, 
and hence the corresponding limit manifolds are of the form (6.7). This 
implies that $\varepsilon$-thick regions converge to hyperbolic 
metrics, giving rise to the $H$ factor in (6.4).

\medskip

  It would be of interest to construct large families of examples of vacuum 
space-times exhibiting the conclusions (and hypotheses) of Theorem 6.2. 

\medskip

{\sf Recent Note}: (January, 04). The recent work of Grisha Perelman [49]-[51], 
currently under evaluation in the mathematics community, implies a solution of 
Thurston's Geometrization Conjecture, and hence in particular the Poincar\'e 
Conjecture.

\bibliographystyle{plain}

\begin{thebibliography}{WWW}
\bibitem [49]{49} G. Perelman, The entropy formula for the Ricci flow and 
its geometric applications, math.DG/0211159.
\end{thebibliography}

\bigskip
\begin{center}
August, 2002
\end{center}

\medskip
\noindent
\address{Department of Mathematics\\
S.U.N.Y. at Stony Brook\\
Stony Brook, N.Y. 11794-3651}

\noindent
\email{anderson@math.sunysb.edu}

\end{document}